%% file: milp_market.tex
\documentclass[prodmode,license]{acmsmall-ec16}

\pdfoutput=1

\usepackage[ruled]{algorithm2e}

\usepackage{amsmath,amsfonts,amssymb,bbm}
\usepackage[numbers,sort&compress]{natbib} 

\SetArgSty{textrm}  
\SetAlFnt{\small}
\SetAlCapFnt{\small}
\SetAlCapNameFnt{\small}
\SetAlCapHSkip{0pt}
\IncMargin{-\parindent}

\newenvironment{mechanism}{\begingroup%
\begin{algorithm}}{\end{algorithm}\endgroup}

\usepackage{mathtools}
\usepackage{xspace}
\usepackage{enumerate}
\usepackage{graphicx}
\usepackage{microtype}

\usepackage{standalone}
\usepackage{mathtools}
\usepackage{tikz}
\usepackage{tikz-qtree}
\usepackage{soul}
\usetikzlibrary{svg.path}
\usetikzlibrary{trees}

\newcount\Comments  
\definecolor{thegreen}{rgb}{0,0.6,0.1}
\definecolor{thered}{rgb}{0.8,0.2,0.1}
\definecolor{teal}{rgb}{0.1,0.6,0.6}
\newcommand{\kibitz}[2]{\ifnum\Comments=1\textcolor{#1}{#2}\fi}

\input{notation}

\begin{document}
\SetEndCharOfAlgoLine{}


\title{Arbitrage-Free Combinatorial Market Making via Integer Programming}

\author{CHRISTIAN KROER
\affil{Carnegie Mellon University}
MIROSLAV DUD\'IK
\affil{Microsoft Research}
S\'EBASTIEN LAHAIE
\affil{Microsoft Research}
SIVARAMAN BALAKRISHNAN
\affil{Carnegie Mellon University}
}

\begin{abstract}
We present a new combinatorial market maker that operates arbitrage-free
combinatorial prediction markets specified by integer programs. Although the problem of arbitrage-free
pricing, while maintaining a bound on the subsidy provided by the market maker, is \#P-hard
in the worst case, we posit that the typical case might be amenable to modern integer programming (IP) solvers.
At the crux of our method is the Frank-Wolfe (conditional gradient) algorithm which is used to implement a Bregman projection aligned with the market maker's cost function, using an IP solver as an oracle.
We demonstrate the tractability and improved accuracy of our approach on real-world
prediction market data from combinatorial bets placed on the 2010 NCAA
Men's Division I Basketball Tournament, where the outcome space is of
size $2^{63}$. To our knowledge, this is the first implementation
and empirical evaluation of an arbitrage-free combinatorial prediction market on this scale.
\end{abstract}



\doi{2940716.2940767}

%


\begin{bottomstuff}

Author addresses:
C. Kroer, Computer Science Dept, CMU; \url{ckroer@cs.cmu.edu}; M. Dud\'ik and S. Lahaie, Microsoft Research; \{mdudik,slahaie\}@microsoft.com; S. Balakrishnan, Dept of Statistics, CMU; \url{siva@stat.cmu.edu}. This work was done while C. Kroer and S. Balakrishnan were at Microsoft Research.
\end{bottomstuff}

\maketitle

\input{intro}
\input{setting}
\input{design}
\input{fwmm}

\input{dataset}
\input{eval}
\input{conclusion}

\begin{small}
\bibliographystyle{ACM-Reference-Format-Journals}
\bibliography{refs_short}
\end{small}
\newpage
\input{appendix}

\end{document}

%% file: notation.tex
\newcommand{\wo}{\backslash}
\newcommand{\inprod}{\cdot}

\newcommand{\Fig}[1]{Fig.~\ref{fig:#1}}
\newcommand{\Sec}[1]{Sec.~\ref{sec:#1}}
\newcommand{\Eq}[1]{Eq.~\eqref{eq:#1}}
\newcommand{\Prop}[1]{Proposition~\ref{prop:#1}}
\newcommand{\App}[1]{Appendix~\ref{app:#1}}
\newcommand{\Mech}[1]{Mechanism~\ref{mech:#1}}
\newcommand{\Algo}[1]{Algorithm~\ref{algo:#1}}

\newcommand{\Ex}[1]{Example~\ref{ex:#1}}

\newcommand{\I}{\mathcal{I}}
\newcommand{\M}{\mathcal{M}}
\newcommand{\tM}{\tilde{\M}}

\newcommand{\X}{\mathcal{X}}
\newcommand{\J}{\mathcal{J}}
\renewcommand{\S}{\mathcal{S}}
\newcommand{\Eset}{\mathcal{E}}
\newcommand{\eps}{\varepsilon}
\newcommand{\C}{\mathcal{C}}
\newcommand{\Vsigma}{V_\sigma}
\newcommand{\ind}{\mathbb{I}}

\newcommand{\A}{\mathbf{A}}
\newcommand{\tA}{\mathbf{\tilde{A}}}
\newcommand{\D}[3][]{D_{#1}(#2\Vert#3)}
\newcommand{\vb}{\mathbf{b}}
\newcommand{\tvb}{\mathbf{\tilde{b}}}
\newcommand{\vc}{\mathbf{c}}
\newcommand{\one}{1}
\newcommand{\vdelta}{\boldsymbol{\delta}}
\newcommand{\vphi}{\boldsymbol{\phi}}
\newcommand{\vtheta}{\boldsymbol{\theta}}
\newcommand{\svtheta}{\boldsymbol{\theta^\star}}
\newcommand{\hvtheta}{\boldsymbol{\hat{\theta}}}

\newcommand{\svdelta}{\boldsymbol{\delta^\star}}
\newcommand{\veta}{\boldsymbol{\eta}}

\newcommand{\vomega}{\boldsymbol{\omega}}
\newcommand{\R}{\mathbb{R}}
\newcommand{\E}{\mathbb{E}}
\renewcommand{\Pr}{\mathbb{P}}
\newcommand{\vzero}{\boldsymbol{0}}
\newcommand{\vmu}{{\boldsymbol{\mu}}}
\newcommand{\svmu}{{\boldsymbol{\mu^\star}}}
\newcommand{\hvmu}{\hat{\vmu}}
\newcommand{\vp}{\boldsymbol{p}}

\newcommand{\vz}{\boldsymbol{z}}
\newcommand{\hvz}{\boldsymbol{\hat{z}}}
\newcommand{\hz}{\hat{z}}
\newcommand{\vu}{{\boldsymbol{u}}}
\newcommand{\Z}{\mathcal{Z}}

\newcommand{\bR}{\bar{R}}
\newcommand{\hsigma}{{\hat{\sigma}}}
\newcommand{\Project}{\texttt{ProjectFW}\xspace}
\newcommand{\InitFW}{\texttt{InitFW}\xspace}

\newcommand{\card}[1]{\lvert#1\rvert}
\newcommand{\Bracks}[1]{\left[#1\right]}
\newcommand{\bigBracks}[1]{\bigl[#1\bigr]}
\newcommand{\BigBracks}[1]{\Bigl[#1\Bigr]}

\newcommand{\parens}[1]{(#1)}
\newcommand{\Parens}[1]{\left(#1\right)}
\newcommand{\bigParens}[1]{\bigl(#1\bigr)}

\newcommand{\set}[1]{\{#1\}}
\newcommand{\braces}[1]{\{#1\}}
\newcommand{\bigSet}[1]{\bigl\{#1\bigr\}}
\newcommand{\Set}[1]{\left\{#1\right\}}

\newcommand{\ltValue}{\textit{lt}}
\newcommand{\eqValue}{\textit{eq}}
\newcommand{\gtValue}{\textit{gt}}

\newcommand{\citeposs}[1]{\citeANP{#1}'s~\citeyearpar{#1}}

\newcommand{\figsqueezeR}{\vspace{-18pt}} 
\newcommand{\figsqueeze}{\vspace{-0pt}} 

\DeclareMathOperator{\conv}{conv}

\DeclareMathOperator{\diam}{diam}
\DeclareMathOperator*{\argmax}{argmax}
\DeclareMathOperator*{\argmin}{argmin}

\newcommand{\Duke}{\textit{Duke}\xspace}
\newcommand{\Cornell}{\textit{Cornell}\xspace}
\newcommand{\Predictalot}{\textit{Predictalot}\xspace}

\newtheorem{assu}[theorem]{Assumption}
\newenvironment{assumption}{%
\italicenvfalse
\begin{assu}}{\end{assu}\italicenvtrue}

%% file: intro.tex
\section{Introduction}
\label{sec:intro}
Prediction markets have been successfully used to elicit and aggregate forecasts
in a variety of domains, including
business~\cite{Spa:03,C07}, politics~\cite{Berg+08}, and entertainment~\cite{PenScience:01}.
In a prediction market, traders buy and sell securities with values that depend on some unknown future outcome.
For instance, a play-money prediction market that Yahoo!\ ran for the 2010 NCAA Men's Division I Basketball Tournament
%
included a security that paid out 1 point if the team from $\Duke$ were to win the 
championship and 0 points otherwise. Thus, when the price of the
security was 0.15, traders who believed that $\Duke$'s probability of winning was larger than 0.15 were incentivized to buy shares of the security, and those that believed it was lower were incentivized to sell. The market price can be interpreted as an aggregate belief and used as a forecast.

We study prediction markets implemented by a centralized algorithm called a
\emph{cost-based market maker}~\cite{ChenPe07,AbernethyChVa11}.
All shares are bought from and sold to the market maker, rather than
between traders, and the market maker uses a convex potential function
to determine current security prices.
Compared with an exchange, which matches buyers and sellers, a market-maker mechanism is particularly desirable in \emph{combinatorial markets}, which offer securities on interrelated propositions. For instance, the NCAA 2010 market included securities on events ``\Duke wins more games than \Cornell'' and ``a team from the \emph{Big East} conference wins the championship'' as well as many others. Because of the large number of securities in combinatorial markets, there may be no sellers interested in trading with a given buyer, a problem known as low liquidity. In contrast, a market maker is always available to trade, thus providing liquidity and allowing incorporation of information in the market.

Designers of cost-based markets aim to meet several desirable properties, including
\emph{boundedness of loss} suffered by the market maker
and \emph{absence of arbitrage}, that is, risk-free profitable trades.
Bounded loss
is a necessity for money markets, otherwise the market operator risks bankruptcy. Lack of arbitrage is also highly desirable.
First, we would like to attract traders that provide information rather than computation. Second, arbitrage-free markets produce more accurate forecasts.
While \citet{AbernethyChVa11} provide a complete theoretical characterization of cost-based markets with bounded loss and no arbitrage, pricing in such markets is NP-hard or \#P-hard for even the simplest combinatorial settings~\citep{ChenEtAl08}.
Previous solutions
restrict the betting language to allow polynomial time
algorithms~\citep[e.g.,][]{Chen:07,Chen:08b,XiaPe11}
or devise approximations~\citep{DudikLaPe12}.

In this paper we move beyond the hardness barrier. We hypothesize that
while the pricing may be difficult in the worst case, a typical case
is amenable to modern integer programming solvers. Guided by this
hypothesis, we propose a fully general, bounded-loss, arbitrage-free
market maker based on integer programming (IP) methods.
%
%
Our market maker is guaranteed to maintain bounded loss, and attempts to remove arbitrage by making calls to the IP solver.


Our mechanism begins with any bounded-loss cost function and adds two ingredients. First, we use an integer program to specify the set of valid payoff vectors, each of which enumerates the security payoffs in a single outcome.
Arbitrage-free prices are exactly the convex combinations of valid payoff vectors. Second, as we run the cost-based market, we periodically remove arbitrage by projecting the market state onto the set of arbitrage-free prices,
using the IP solver as an oracle within the projection algorithm. The integer program for payoff vectors is typically compact in size and easy to specify based on security definitions.
For instance, when securities correspond to logical propositions, outcomes correspond to truth-assignment of literals, and each valid payoff vector enumerates 0 for false and 1 for true propositions in a given outcome. Conjunctions and disjunctions are easily expressed within an IP, so we have a compact representation for all problems in NP.
To implement projection, we use the Frank-Wolfe algorithm~\cite{FrankWo56,Jaggi13},
also known as the conditional gradient algorithm, which is well-suited to our setting because it only accesses the target set (in our case, the set of valid payoff vectors) through the operation of linear optimization, which can be handled by an IP solver. The projection that we consider is the \emph{Bregman projection}, which generalizes the Euclidean projection to arbitrary convex potentials.

There are two specific issues in applying the Frank-Wolfe (FW) algorithm within cost-based markets. First, while all iterates of the FW algorithm are within the convex hull of valid payoff vectors, and therefore arbitrage-free, we need to ensure that bounded loss is maintained. In \Sec{stopping}, we show how to achieve this by a suitable modification of the stopping condition of FW. The second, seemingly more serious concern, is that the projection problems that arise for common cost functions, such as Hanson's \citeyear{Hanson03,Hanson07} logarithmic
market scoring rule (LMSR), exhibit derivatives that go to infinity at the border of the set of arbitrage-free prices, which violates the assumptions of the FW algorithm.
Fortunately, we can adapt a recently developed variant of FW~\cite{KrishnanLaSo15}, designed for the case when the derivative might grow to infinity, but its growth is suitably controlled, which is the case for LMSR.

Our approach, which we call the \emph{Frank-Wolfe market maker} (FWMM), is related to \citeposs{DudikLaPe12} linearly-constrained market maker (LCMM),
which also alternates trades and (partial) arbitrage removal. While FWMM uses linear constraints in the IP to define valid payoff vectors, the arbitrage removal in LCMM is driven by a set of linear constraints on the arbitrage-free prices (i.e., the convex hull of valid payoff vectors).
The IP constraints of FWMM can be used directly in LCMM, as linear-programming relaxations, but they are usually too loose,
so tighter constraints
%
%
need to be derived \textit{ad hoc} for each new security type,
sometimes using involved combinatorial reasoning~\cite{DudikLaPe12,DudikEtAl13}.
Since LCMM updates are usually substantially faster than solving an IP, the arbitrage-removal steps of LCMM and FWMM can be interleaved,
and
the more expensive
projection step of FWMM should be invoked only after LCMM cannot remove much arbitrage.

We evaluate the efficacy
of FWMM on Yahoo!'s NCAA 2010 basketball tournament prediction market data, from which we extracted 88k trades on 5k securities in a combinatorial market with $2^{63}$ outcomes. Once the projections become practically fast,
FWMM achieves superior accuracy to LCMM. Our experiments also show that the initial phase of the projection algorithm, which involves calls to the IP solver to decide which securities can be logically settled given the games completed so far, is fast even for the largest problem sizes. The results from this initial phase can be propagated as a partial outcome into the cost function, which yields an improvement over LCMM even when the overall projection algorithm is too slow.

Tournaments have previously been considered by \citet{Chen:08b} and \citet{XiaPe11}. Both focus on restricted (but non-trivial) tournament betting languages that yield tractability, but cannot, for instance, handle comparisons. In contrast, our approach works for general outcome spaces that can be represented by an IP, rather than only tournaments. Our work is closely related to the applications of Frank-Wolfe and integer programming to inference in graphical models~\cite{KrishnanLaSo15,BelangerShMc13}, but needs to address several issues specific to incentives and information revelation in prediction markets.

%% file: setting.tex
\section{Preliminaries}
\label{sec:prelims}

We begin with an overview of cost-based market making~\citep{ChenPe07,AbernethyChVa11}
and then provide a high-level outline of our approach.
As a running example we use the NCAA 2010 Tournament:
a single-elimination tournament with
64 teams playing over 6 rounds, meaning that in
each round, half of the remaining teams are eliminated.

\subsection{Cost-based market making}

Let $\Omega$ denote a finite set of
\emph{outcomes}, corresponding to mutually exclusive and
exhaustive states of the world.
We are interested in eliciting expectations
of binary random variables $\phi_i:\Omega\to\{0,1\}$, indexed by $i\in\I$, which
model the occurrence of various events such as ``\Duke wins the NCAA championship.''
%
%
Each variable $\phi_i$ is associated with a \emph{security}, which is
a contract that pays out $\phi_i(\omega)$ dollars when the outcome
$\omega$ occurs. Therefore, the random variable $\phi_i$ is also
called the \emph{payoff function}. Binary securities pay out \$1 if
the specified event occurs and \$0 otherwise. The vector
$(\phi_i)_{i\in\I}$ is denoted~$\vphi$. Traders buy \emph{bundles}
$\vdelta\in\R^\I$ of security shares issued by a central market maker;
negative entries in $\vdelta$ are permitted and correspond to
short-selling. A trader holding a bundle $\vdelta$ receives a
(possibly negative) payoff $\vdelta\inprod\vphi(\omega)$ when
$\omega\in\Omega$ occurs.


Following \citet{ChenPe07} and \citet{AbernethyChVa11}, we assume that
the market maker determines security prices using a convex and differentiable
potential function
$C:\R^\I\to\R$ called a \emph{cost function}.
The state of the market is specified by a vector $\vtheta\in\R^\I$
listing the number of shares of each security sold so far by
the market maker. A trader wishing to buy a bundle $\vdelta$ in the market state $\vtheta$
must pay $C(\vtheta+\vdelta)-C(\vtheta)$ to the
market maker, after which the new state becomes $\vtheta+\vdelta$.
Thus, the vector of instantaneous prices in the state $\vtheta$ is
$\vp(\vtheta)\coloneqq\nabla C(\vtheta)$. Its entries can be interpreted
as market estimates of $\E[\phi_i]$:
a trader can make
an expected profit by
buying (at least a small amount) of
the security $i$ if she believes that $\E[\phi_i]$ is
larger than the instantaneous price
$p_i(\vtheta)=\partial C(\vtheta)/\partial\theta_i$, and by selling if she believes
that $\E[\phi_i]$ is lower than $p_i(\vtheta)$; therefore, risk neutral traders with sufficient
budgets maximize their expected profits by moving the price vector to match their
expectation of $\vphi$.

\begin{example}\emph{Logarithmic market-scoring rule (LMSR).}
\label{ex:lmsr}
Hanson's \citeyear{Hanson03,Hanson07} logarithmic
market scoring rule (LMSR) is a cost function for a \emph{complete market}.
In a complete market, $\I=\Omega$ and securities are
indicators of individual
outcomes, $\phi_i(\omega)=\one\set{\omega=i}$, where $\one\set{\cdot}$ denotes the binary indicator, equal to $1$ if true and $0$ if false. Thus,
traders can express arbitrary probability distributions
over $\Omega$.
For instance,
to set up a complete market for the number of wins of \Duke in the six-round NCAA tournament,
we would set $\I=\Omega=\set{0,1,\dotsc,6}$. LMSR
has the form $C(\vtheta)=\log\parens{\sum_{i\in\I} e^{\theta_i}}$ and
prices $p_i(\vtheta)=e^{\theta_i}/\parens{\sum_{j\in\I} e^{\theta_j}}$.
\end{example}

\begin{example}%
\label{ex:sum}%
\emph{Sum of independent markets.}
Now consider a market with 7 securities for the number of wins of \Duke
and an additional 7 securities for the number of wins of \Cornell.
The outcome space consists of pairs of numbers between 0 and 6, but not all pairs
are possible, because if \Duke and \Cornell win rounds~1--4, they meet in round~5
and only one advances. Thus,
$\Omega=\set{(\omega_1,\omega_2)\in\set{0,\dotsc,6}^2:\:\min\set{\omega_1,\omega_2}\le 4}$.
Securities are indexed by pairs $\I=\set{1,2}\times\set{0,\dotsc,6}$, with the first
entry indicating the school and the second the number of wins, yielding the payoff
functions $\phi_{j,x}(\vomega)=\one\set{\omega_j=x}$. A natural
cost function is the sum of LMSRs, $C(\vtheta)=\smash[b]{\sum_{j=1}^2\log\parens{\sum_{x=0}^6 e^{\theta_{j,x}}}}$, which yields prices $p_{j,x}(\vtheta)=e^{\theta_{j,x}}/\parens{\sum_{y=0}^6 e^{\theta_{j,y}}}$. Thus, prices vary independently for each school, as if we ran two
separate markets.
\end{example}

\subsection{Arbitrage, marginal polytope and Bregman projection}

We consider two standard desiderata for cost-based markets. The first
is the \emph{bounded loss} property: there should be a constant which bounds the ultimate loss of the market maker once the outcome is determined, regardless of how many shares of each security are sold. The second is the \emph{no arbitrage} property: there should be no
trade that guarantees a positive profit, regardless of the outcome.
%
%
Following \citet{AbernethyChVa11}, we next relate bounded loss to properties of the convex conjugate of $C$, and review equivalence between optimal arbitrage removal and
\emph{Bregman projection}.

Given a cost function $C$, let $R$ denote its \emph{convex conjugate},
\begin{equation}
\label{eq:def:R}
  R(\vmu)\coloneqq\sup_{\vtheta'\in\R^\I}\bigBracks{\vtheta'\inprod\vmu-C(\vtheta')}
\enspace,
\end{equation}
which is itself a convex function on $\R^\I$, allowed to take on the
value $\infty$. If the market is in a state $\vtheta=\vzero$ and a
trader believes that $\E[\vphi]=\vmu$, then her expected profit for the bundle $\vtheta'$ is $\vtheta'\inprod\vmu-\bigParens{C(\vtheta')-C(\vzero)}$, which is maximized by \Eq{def:R},
omitting the constant term $C(\vzero)$. More
generally, the maximum expected profit of a trader with a belief
$\vmu$ in a market state $\vtheta$ can be shown to equal the
\emph{mixed Bregman divergence}, defined as
\[
  \D{\vmu}{\vtheta}\coloneqq R(\vmu) + C(\vtheta) - \vtheta\inprod\vmu
\enspace.
\]
Convex conjugacy implies that $\D{\vmu}{\vtheta}\ge 0$, with equality if and only if
$\vmu=\vp(\vtheta)$, which is equivalent to $\vtheta\in\partial R(\vmu)$, where $\partial R$ is the subdifferential of $R$.

\begin{example}
  For the LMSR, $R(\vmu)$ is equal to negative entropy whenever $\vmu$ is a probability
  distribution and $\infty$ otherwise, i.e., $R(\vmu)=
      \ind\set{\vmu\in\Delta}+\sum_{i\in \I} \mu_{i}\ln \mu_{i}$, where $\Delta$ is
       the set of probability distributions on $\Omega$ and $\ind\set{\cdot}$ denotes
       the convex indicator, equal to $0$ if true and $\infty$ if false.
Bregman divergence is the Kullback-Leibler (KL) divergence,
$\D{\vmu}{\vtheta} = \ind\set{\vmu\in\Delta} +
      \sum_{i\in \I} \mu_{i}\ln\Parens{\mu_{i} / p_{i}(\vtheta)}$,
which is an information-theoretic measure of the difference
between two probability distributions.
\end{example}

Let $\Z\coloneqq\set{\vphi(\omega):\:\omega\in\Omega}$ denote the (finite) set
of all valid payoff vectors, and $\M$ be its convex hull, called the
\emph{marginal polytope}. The marginal polytope is exactly the set of vectors
$\vmu$ that can be written as expectations $\E[\vphi]$ under some
probability distribution over~$\Omega$, so we refer to elements of
$\M$ as \emph{coherent beliefs} or \emph{coherent prices}.
\citet{AbernethyChVa11} show that a cost-based market maker has the bounded loss property if and only if $\max_{\vz\in\Z} R(\vz)<\infty$. We assume that this is the case for the conjugate of our cost~$C$. Note that this assumption is satisfied for LMSR, because negative entropy equals zero at the vertices of the simplex. It is also satisfied in Example~\ref{ex:sum}, where $R(\vmu)$ is the sum of negative entropies of the two markets.

%
%

Given a state $\vtheta$, we define the
\emph{Bregman projection} of $\vtheta$ on $\M$ as the point
\[
  \svmu\coloneqq\argmin_{\vmu\in\M} \D{\vmu}{\vtheta}
\enspace.
\]
The Bregman projection is related to an optimal arbitraging trade by the following standard result (the proof is in \App{proof:arb:proj} for completeness):

\begin{proposition}
If the market is in a state $\vtheta$, the guaranteed profit of any trader is at most $\D{\svmu}{\vtheta}$ where $\svmu$ is the Bregman projection of $\vtheta$ on $\M$. Furthermore, this profit is achieved by any trade $\svdelta$ moving the market to a state $\svtheta$ with $\vp(\svtheta)=\svmu$.
\label{prop:arb:proj}
\end{proposition}

This means that an arbitrage opportunity exists whenever the prices are incoherent, since $\vp(\vtheta)\not\in\M$ implies that $\D{\svmu}{\vtheta}>0$. After the trade $\svdelta$, we have $\vp(\svtheta)=\svmu\in\M$ and thus there is no arbitrage opportunity in the market.

\subsection{The outline of Frank-Wolfe market maker (FWMM)}
\label{sec:fwmm:outline}

The mechanism proposed in this paper, called Frank-Wolfe market maker, alternates between processing trades according to the cost~$C$ and removing arbitrage. In the arbitrage removal step, our goal is to find the state $\svtheta$ from \Prop{arb:proj}. We do this by solving the Bregman projection problem using the Frank-Wolfe (FW) algorithm, which reduces the Bregman projection problem to a sequence of linear programs of the form
\[
  \min_{\vmu\in\M} \vc\inprod\vmu
\enspace,
\]
for suitably chosen vectors $\vc$.
Since the optimum of a linear program occurs at a vertex, reducing the Bregman projection problem to a sequence of linear programs results in an important simplification. Instead of specifying the marginal polytope $\M$, whose description can be exponentially large in the number of securities, it suffices to describe its vertices $\Z$, which we show can be done via a compact set of linear inequalities together with integer constraints. More precisely,
we assume
that the set $\Z$ is described by a matrix $\A$ and a vector $\vb$ such that
\begin{equation}
\label{eq:Z}
   \Z = \Set{ \vz\in\set{0,1}^\I:\: \A^\top\vz\ge\vb }
\enspace.
\end{equation}
Viewed in this way, the FW algorithm solves the Bregman projection problem by solving a sequence of integer programs. We refer to the linear constraints describing the set $\Z$ as \emph{IP constraints}.


\begin{example}
We next derive IP constraints for the market
for the number
of wins of $\Duke$ and $\Cornell$ from Example~\ref{ex:sum}. First,
there are exclusivity and exhaustivity constraints of the form $\sum_{x=0}^6 z_{j,x}=1$ for $j\in\set{1,2}$, corresponding to the fact that in any outcome $\vomega$, for each $j$, exactly one of the securities $\phi_{j,x}(\vomega)$ will equal 1 across $x\in\set{0,\dotsc,6}$. However, these two constraints do not capture the fact that at most one of the teams can have exactly 5
or 6 wins. Specifically, in any outcome $\vomega$, we have
\[
 \phi_{1,5}(\vomega)+\phi_{2,5}(\vomega)
 +
 \phi_{1,6}(\vomega)+\phi_{2,6}(\vomega)
 \le 1
\enspace.
\]
Thus, we also include the third constraint:
$
 z_{1,5}+z_{2,5}+z_{1,6}+z_{2,6}\le 1
$.
Our reasoning so far shows that any valid payoff vector satisfies the three mentioned constraints. It can be verified that any vector $\vz$ satisfying these constraints is valid, i.e., it corresponds to $\vphi(\vomega)$ for some $\vomega\in\Omega$,
so these three constraints correctly specify $\Z$.
\end{example}



\subsection{Linearly-constrained market maker (LCMM)}
\label{sec:LCMM}

The FW algorithm relies on the ability to solve integer programs
(IPs), which can take exponential time in the worst case. Therefore,
our mechanism also incorporates fast (poly-time) partial arbitrage removal
similar to \citeposs{DudikLaPe12} linearly-constrained market maker
(LCMM).

In LCMM, arbitrage is partly removed by considering
a set of linear constraints that must be satisfied by coherent prices.
Namely, an LCMM takes as an input a relaxation $\tM\supseteq\M$
described by linear constraints called \emph{LCMM constraints}:
\[
   \tM=\set{\vmu\in\R^\I:\:\tA^\top\vmu\ge\tvb}
\enspace.
\]
When any LCMM constraint is violated, there is an arbitrage
opportunity in the market, with an easy-to-compute arbitraging trade. LCMM acts as an arbitrager
until none of the constraints are
violated.
Since $\tM$ is a relaxation of $\M$, the resulting state is not necessarily arbitrage-free.

Assuming we have a description of $\Z$ using IP constraints specified by a matrix $\A$ and a vector $\vb$, one simple strategy is to construct $\tM$ as a linear-program (LP) relaxation of~$\Z$, i.e.,
\begin{equation}
\label{eq:tM:simple}
   \tM=\set{\vmu\in\R^\I:\:\text{$\mu_i\in[0,1]$ for all $i\in\I$ and $\A^\top\vmu\ge\vb$}}
\enspace.
\end{equation}
These constraints are satisfied by all $\vz\in\Z$ and hence also by their convex combinations $\vmu\in\M$. Generally, this relaxation is only a loose superset of $\M$, so various \textit{ad hoc} strategies are required to obtain a tighter $\tM$~\citep{DudikLaPe12,DudikEtAl13}. We present one example of such a strategy in \Sec{design}, for the class of comparison securities.

%% file: design.tex
\section{Market Design}
\label{sec:design}

We next show how to instantiate the market design elements of \Sec{prelims} in real-world combinatorial markets, including the NCAA 2010 tournament
evaluated in \Sec{exp}. Namely, we need to define: (i) the payoff function~$\vphi$, (ii) the cost function~$C$, (iii) the initial market state $\vtheta$, (iv) the IP constraints describing $\Z$, and (v) the LCMM constraints describing~$\tM$.
We also need to consider how the cost and market state should be updated as the true outcome is gradually revealed over time.
For example, in the NCAA tournament, 63 games
play out over the course of several weeks and we would like to fix prices of securities whose payoff has  already been determined.

\subsection{Compositional market design}
\label{sec:compositional}

We use a compositional market design
along the lines of \citet{DudikEtAl13}, which is a generalization of
the sum of LMSRs structure of Example~\ref{ex:sum}.
The market construction begins with a collection of random variables $X_j:\Omega\to\X_j$, indexed by $j\in\J$, whose marginal distributions we wish to elicit, such as the number of wins of \Duke and \Cornell in Example~\ref{ex:sum}.
Securities are indexed by $i=(j,x)$, with $j\in\J$ and $x\in\X_j$,
and correspond to indicators of the events $X_j=x$, i.e.,
\[
   \phi_{j,x}(\omega)=\one\set{X_j(\omega)=x}
\enspace.
\]
The cost function is the sum of LMSRs across the random variables $X_j$:
\begin{equation}\label{eq:lmsr-cost}
\textstyle
    C(\vtheta) = b\sum_{j\in\J}\ln\Parens{\sum_{x\in\X_j} e^{\theta_{j,x}/b}}
\enspace,
\end{equation}
where $b>0$ is the
\emph{liquidity parameter} controlling how fast the prices
change in response to trading. A smaller value of $b$ (lower liquidity) means
prices rise faster as shares are purchased; a larger value of $b$ (higher liquidity) yields slower changes. As in \Ex{sum}, \Eq{lmsr-cost} implies that we effectively run an independent LMSR market for each $X_j$. Thus, in the absence of arbitrage removal steps, we say that $C$ implements the \emph{independent markets} cost function.


Initially, our market contains no random variables and hence no securities. The market operator
can create new random variables and specify their relationship to any existing
variables. At the time of creation of a new variable $X_j$, the operator specifies (i) its domain $\X_j$, (ii) the mapping $X_j(\omega)$, (iii) initial prices $\mu_{j,x}$ across $x\in\X_j$
(these prices determine the initial-state coordinates $\theta_{j,x}$),
(iv) IP constraints to restrict $z_{j,x}$
across $x\in\X_j$,
and (v) LCMM constraints to restrict $\mu_{j,x}$
across $x\in\X_j$.
Due to the additive structure of the cost $C$, new variables $X_j$ can be added at any time during the run of the market without affecting prices of existing securities.

Below we specify the items (i)--(v) for different types of random variables in our market. When describing the IP constraints on $\vz$ and LCMM constraints on $\vmu$, we use the notation $z\braces{X_j=x}$ and $\mu\braces{X_j=x}$ for the entries $z_{j,x}$ and $\mu_{j,x}$, respectively. We also allow random variables with names other than $X_j$, e.g., $X$ or $G_{r,t}$, and use the notation such as $z\braces{X=x}$ and $\mu\braces{X=x}$ for the corresponding entries of $\vz$ and $\vmu$.

When adding a new random variable $X$, the initial prices $\mu\braces{X=x}$ can be chosen based on the prices of the random variables present in the market.
New IP constraints always include the exclusivity and exhaustivity constraint,
$\sum_{x\in\X} z\braces{X=x}=1$, but additional constraints may be needed to correctly describe the mapping~$X(\omega)$. We add LCMM constraints using the simple strategy mentioned in \Sec{LCMM}, as an LP relaxation of IP constraints, with an exception of one variable type (comparison variables).

Our market contains random variables of the following types:

\paragraph{Atomic tournament variables}

These random variables model outcomes in a single-elimination tournament with $k$ rounds
and $2^k$ teams. Teams are numbered $1$ through~$2^k$. In the first round, there are $2^{k-1}$ games, between teams $2i-1$ and $2i$, and the resulting $2^{k-1}$ winners advance to the second round, where again teams are matched in the order of increasing indices and the winners advance to the next round etc. The team $t$ is associated with the random variable $X_t$ whose outcome is the total number of wins of team $t$, i.e., $\X_t=\set{0,\dotsc,k}$.

\begin{figure}[t]
  \begin{center}
    \scalebox{0.95}{
      \input{figures/tournament_tree}
    }
  \end{center}
\figsqueeze%
  \caption{An example of a tournament with four teams. The domains of the game outcome variables $G_{r,t}$ are shown on the right. The shown variables are equivalent to additional game variables: $G_{1,1}\equiv G_{1,2}$, $G_{1,3}\equiv G_{1,4}$, and
  $G_{2,1}\equiv G_{2,2}\equiv G_{2,3}\equiv G_{2,4}$.}
  \label{fig:tournament_example}
\figsqueeze%
\end{figure}
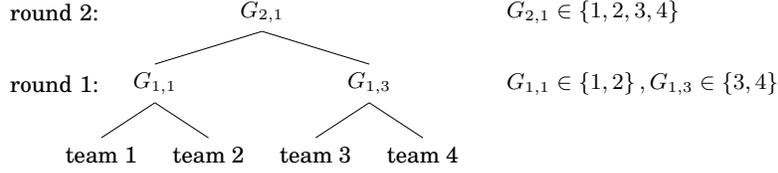

We also have random variables
corresponding to the games played, with the outcome of each variable being
the winner of the corresponding game.
For a team $t$ and round~$r$, let $G_{r,t}$ denote the game that the team $t$
will play in the $r$-th round if it advances to that point. We are slightly abusing notation, because $G_{r,t}$ and $G_{r,t'}$ can refer to the same game (and hence the same random variable) for distinct $t$ and $t'$ (see \Fig{tournament_example}). For instance $G_{k,t}\equiv G_{k,t'}$ for all $t,t'$, as there is only one game (the finals) in round $k$.
With this notation in hand, we can introduce the IP constraints relating the entries of $\vz$ representing game and team variables:
\begin{align*}
  z\set{X_t=r}&=z\set{G_{r,t}=t}-z\set{G_{r+1,t}=t}
&&\text{for all $t$ and $r<k$,}
\\
  z\set{X_t=k}&=z\set{G_{k,t}=t}
&&\text{for all $t$.}
\end{align*}
%
%
LCMM constraints are just LP relaxations of the above, i.e., they are the same as the IP constraints, with $z\braces{\cdot}$ replaced with $\mu\braces{\cdot}$.
%
%
The market operator needs to specify initial prices $\mu\set{X_t=r}$ and \mbox{$\mu\set{G_{k,t}=t}$} explicitly, based for instance on the past performance of teams.

\paragraph{Sums}
Given a set of existing random variables $X_1,\dotsc,X_n$ taking on integer values with the minimum and maximum values $m_j\coloneqq\min\X_j$ and $M_j\coloneqq\max\X_j$, we define a new random variable $X$ to represent their sum,
\[
   X(\omega) \coloneqq X_1(\omega)+\dots+X_n(\omega)
\enspace,
\]
with the domain $\X=\set{m,m+1,\dotsc,M}$ where $m=\sum_{j=1}^n m_j$ and $M=\sum_{j=1}^n M_j$. The initial prices are set proportional to a discretized Gaussian distribution with the mean and variance equal to the sum of means and variances of $X_1$ through $X_n$, under the distribution described by the current prices $\mu\braces{X_j=x}$.

We introduce the following IP constraint:
%
\[
\textstyle
   \sum_{x\in\X} x\cdot z\set{X=x}
   =
   \sum_{j=1}^n\sum_{x_j\in\X_j} x_j\cdot z\set{X_j=x_j}
\enspace.
\]
As before, the added LCMM constraint is an LP relaxation of the added IP constraint.

\paragraph{Comparisons}
Given two existing random variables $X_1$ and $X_2$ taking on
integer values with the minimum and maximum values $m_j\coloneqq\min\X_j$ and $M_j\coloneqq\max\X_j$, we define a new random variable $X$ with the domain $\set{\ltValue,\eqValue,\gtValue}$ to represent the result of their comparison:
\[
  X(\omega)
\coloneqq
  \begin{cases}
    \ltValue&\text{if $X_1(\omega)<X_2(\omega)$,}
\\  \eqValue&\text{if $X_1(\omega)=X_2(\omega)$,}
\\  \gtValue&\text{if $X_1(\omega)>X_2(\omega)$.}
  \end{cases}
\]
The initialization prices are determined by first considering an integer-valued variable $Y=X_2-X_1$, and initializing its distribution to the discrete Gaussian with the mean equal to the difference of means and the variance initialized to the sum of variances of $X_2$ and $X_1$ under current prices. The initial prices of $X=\ltValue$, $X=\eqValue$ and $X=\gtValue$ are
obtained as
probabilities that $Y<0$, $Y=0$ and $Y>0$. The variable $Y$ is discarded and is not part of the market.

The IP constraints for the new entries of $\vz$ are based on the following four identities:
\begin{align*}
  X_1-X_2&\ge (m_1-M_2)\one\braces{X_1<X_2}
\enspace,\quad
&
  X_1-X_2-1&\ge (m_1-M_2-1)\one\braces{X_1\le X_2}
\enspace,
\\
  X_1-X_2&\le (M_1-m_2)\one\braces{X_1>X_2}
\enspace,\quad
&
  X_1-X_2+1&\le (M_1-m_2+1)\one\braces{X_1\ge X_2}
\enspace.
\end{align*}
To obtain IP constraints, we replace each $X_j$ with
$\sum_{x\in\X_j} x\cdot z\set{X_j\!=x}$ on the left-hand side, and replace
the comparison indicators on the right-hand side by $z\set{X\!=\ltValue}$ for $\one\braces{X_1<X_2}$, and
$z\set{X\!=\ltValue}+z\set{X\!=\eqValue}$ for $\one\braces{X_1\le X_2}$, and similarly for $X_1>X_2$ and $X_1\ge X_2$.

LCMM constraints in this case are not simply an LP relaxation of IP constraints, but instead they yield a tighter set $\tM$. They are based on the following identities, which can be derived from the transitivity of the comparison and the union bound:
\begin{align*}
  \Pr\set{X_1\le x} &\le \Pr\set{X_1 < X_2} + \Pr\set{X_2\le x}
&\text{for all $x\ge m_1$ and $x\le M_2$,}
\\
  \Pr\set{X_1\le x} &\le \Pr\set{X_1 \le X_2} + \Pr\set{X_2< x}
&\text{for all $x\ge m_1$ and $x\le M_2$.}
\end{align*}
For instance, the first inequality follows because $X_1\le x$
implies that either $X_1<X_2$ or $X_2\le x$. Otherwise we would have a contradiction: $X_1\ge X_2>x$. The resulting LCMM constraints are
\begin{align*}
  \mu\set{X_1\le x} &\le \mu\set{X=\ltValue} + \mu\set{X_2\le x}
&\text{for all
$m_1\le x\le M_2$,}
\\
  \mu\set{X_1\le x} &\le \mu\bigSet{X\in\set{\ltValue,\eqValue}} + \mu\set{X_2<x}
&\text{for all
$m_1\le x\le M_2$,}
\end{align*}
with analogous constraints with $X_1$ and $X_2$ swapped (and $\gtValue$
swapped for $\ltValue$). We use the shorthand $\mu\set{X\in\Eset}$ for $\sum_{x\in\Eset}\mu\set{X=x}$.

\subsection{Partial outcomes}
\label{sec:partial}

In a typical combinatorial market, outcomes are gradually revealed
over time.
For example, in the NCAA tournament, 63 games
play out over the course of several weeks. Thus, the market evolves through
a sequence of \emph{partial outcomes} defined as follows:

\begin{definition}
\label{def:partial}
A subset $\sigma\subseteq\I\times\set{0,1}$ is called a \emph{partial outcome}
if there exists a valid payoff vector $\vz\in\Z$ such that $z_i=b$ for all $(i,b)\in\sigma$.
\end{definition}

We write $\I_\sigma\coloneqq\set{i:\:(i,b)\in\sigma \text{ for some }b}$
for the set of securities whose payoffs have been determined, or \emph{settled}, by $\sigma$.
As securities get settled, we would like to fix their prices to 0 or 1.
This is not possible by simply
updating the state, but instead we need to switch to a different cost function while
maintaining the information state of the market. We adapt the construction of \citet{DudikFrWo14}
to our setting.

First, we say
that a vector $\vu\in\R^\I$ is \emph{compatible} with $\sigma$
if $u_i=b$ for all $(i,b)\in\sigma$. We
write $\Vsigma$ for the set of vectors compatible
with $\sigma$---note that $\Vsigma$ is an axis-aligned
affine space of dimension $\card{\I\wo\I_\sigma}$.
Given a partial outcome $\sigma$, we define the set of associated
valid payoffs $\Z_\sigma\coloneqq\Z\cap\Vsigma$, and the associated marginal
polytope $\M_\sigma\coloneqq\conv(\Z_\sigma)$.
We assume that given a partial outcome $\sigma$, the market maker uses
the cost function
\begin{equation}
\label{eq:Csigma}
\textstyle
  C_\sigma(\vtheta)
  =
  \sup_{\vmu\in\Vsigma} \Bracks{\vtheta\inprod\vmu-R(\vmu)}
\enspace,
\end{equation}
whose conjugate is, by definition,
$R_\sigma(\vmu)=R(\vmu)+\ind\set{\vmu\in\Vsigma}$, which coincides with $R$ on
$\M_\sigma$.
The corresponding price map
and Bregman divergence are denoted $\vp_\sigma$ and $D_\sigma$.
The transformation of $C$ to $C_\sigma$ maintains the loss bound of the original market maker (see \App{partial}) and also maintains the information state of the market analogously to conditioning, as our next
example shows.

\begin{example}\emph{Partially settled LMSR.}
Recall that in a complete market,
$\I=\Omega$ and payoff vectors $\vphi(\omega)$ have exactly one entry equal to 1: the entry corresponding to the realized outcome. Therefore,
the partial outcome $\sigma$ can have at most one security settled to 1. If there is such a security $i^\star$ then
the market is fully settled and, by \Eq{Csigma}, we obtain
$C_\sigma(\vtheta)=\theta_{i^\star}$,
$p_{\sigma,i}(\vtheta)=\one\set{i=i^\star}$.
If $\sigma$ only contains securities settled to zero, i.e., the corresponding outcomes have been excluded, the cost function obtained by \Eq{Csigma}
is an LMSR over the remaining outcomes,
$C_\sigma(\vtheta)=\log\parens{\sum_{i\not\in\I_\sigma} e^{\theta_i}}$. The prices
are
$p_{\sigma,i}(\vtheta)=0$ for $i\in\I_\sigma$ and $p_{\sigma,i}(\vtheta)=e^{\theta_i}/\parens{\sum_{j\not\in\I_\sigma} e^{\theta_j}}$ for $i\not\in\I_\sigma$,
so the probability distribution over $\Omega$ described by $\vp_\sigma(\vtheta)$
corresponds to $\vp(\vtheta)$ conditioned on the event
$\omega\not\in\I_\sigma$.
\end{example}

%% file: figures/tournament_tree.tex
\begin{tikzpicture}[level distance=1cm]
  \tikzstyle{every node}=[font=\small]
  \tikzstyle{level 1}=[sibling distance=30mm]
  \tikzstyle{level 2}=[sibling distance=15mm]
  \node(r2g1) {$G_{2,1}$} {
          child{node(r1g1){$G_{1,1}$}{}
              child{node(t1){team 1}
              }
              child{node(t2){team 2}
              }
          }
          child{node(r1g1){$G_{1,3}$}{}
              child{node(t3){team 3}
              }
              child{node(t4){team 4}
              }
          }
  };
  \node[text width=1.4cm, left] at (-2.0,0)
    {round 2:};
  \node[text width=1.4cm, left] at (-2.0,-1)
    {round 1:};
  \node[text width=3.95cm, right] at (3.3,0)
    {$G_{2,1}\in \left\{ 1,2,3,4 \right\}$};
  \node[text width=3.95cm, right] at (3.3,-1)
    {$G_{1,1}\in \left\{ 1,2 \right\}, G_{1,3}\in \left\{ 3,4 \right\}$};
\end{tikzpicture}

%% file: fwmm.tex
\section{Frank-Wolfe Market Maker}
\label{sec:fwmm}

\begin{mechanism}[t]
\caption{Frank-Wolfe Market Maker (FWMM)}
\label{mech:FWMM}
\begin{tabbing}
\textbf{Input:\quad}%
  cost function $C$, initial state $\vtheta_0$, initial partial outcome $\sigma_0$,\\
\hphantom{\textbf{Input:\quad}}%
  LCMM constraints specified by $\tA$, $\tvb$,\\
\hphantom{\textbf{Input:\quad}}%
  IP constraints specified by $\A$, $\vb$,\\
\hphantom{\textbf{Input:\quad}}%
  FW algorithm parameters $\alpha\in(0,1)$, $\eps_0\in(0,1)$, $\eps_D>0$\\[4pt]
  Initialize the market state and partial outcome:
     $\vtheta\gets\vtheta_0,\,\sigma\gets\sigma_0$\\[4pt]
  For $t=1,\dotsc,T$ (where $T$ is an a priori unknown number of trades):\\[4pt]
\hphantom{---}\=
    receive a request for a bundle $\vdelta_t$\\
\>
    sell the bundle $\vdelta_t$ for the cost $C_{\sigma}(\vtheta+\vdelta_t)-C_{\sigma}(\vtheta)$\\
\>
    $\vtheta \gets \vtheta+\vdelta_t$\\
\>
    $\sigma\gets\sigma\cup\set{\text{newly settled securities if any}}$\\[4pt]
\>
    \emph{perform an LCMM step:}\\
\>\hphantom{---}\=
        choose $\veta\ge 0$ such that $C_{\sigma}(\vtheta+\tA\veta)-C_{\sigma}(\vtheta)\le \tvb\inprod\veta$\\
\>\>
        $\vtheta \gets \vtheta+\tA\veta$\\[4pt]
\>
    \emph{perform a projection step:}\\
\>\>
        $(\sigma,\vtheta) \gets \Project(\vtheta;\,C,\sigma,\A,\vb,\alpha,\eps_0,\eps_D)$\\[4pt]
  Observe $\omega$, consistent with $\sigma$\\
  Pay traders
  $\vdelta_1\inprod\vphi(\omega),\,
   \vdelta_2\inprod\vphi(\omega),\,\dotsc,\,
   \vdelta_T\inprod\vphi(\omega)$\\[-16pt]
\end{tabbing}
\end{mechanism}

In this section we fully describe and analyze the Frank-Wolfe market maker (FWMM) outlined in \Sec{fwmm:outline}.

At a high level,
FWMM interleaves rapid pricing according to $C$ with arbitrage removal, while also updating the partial outcome---see \Mech{FWMM}.
There are two kinds of arbitrage removal: fast but only partial arbitrage removal via an LCMM step, and a complete removal of the remaining arbitrage via Bregman projection.
For LCMM steps we use the fast algorithm of \citet{DudikLaPe12}. Bregman projection is implemented via a variant of the Frank-Wolfe (FW) algorithm, which we refer to as \Project and describe later in this section. \Project does not only return a new state $\vtheta$ such that $\vp_\sigma(\vtheta)$ is the Bregman projection of the previous state on $\M_\sigma$. It also
extends the partial outcome to securities that can be logically settled based on all other settled securities.
This permanently removes the specific arbitrage opportunities associated with such securities since their prices become fixed to~0 or~1.


Both arbitrage-removal steps correspond to trades that yield a non-negative profit regardless of the outcome, which means that the loss bound of the original cost $C$ is only improved by the value of this profit. The non-negative profit of LCMM steps follows from \citet{DudikLaPe12}. For \Project, which is an iterative algorithm, we guarantee non-negative profit by designing a suitable stopping condition.


As we mention earlier, while we hope that the IPs created during the run of the FW algorithm are easy to solve, they are NP-hard in general, and so the IP solver can get stuck in a brute-force search. Therefore, we need the ability to interrupt the projection step, for instance, when a new trade arrives. When our implementation, \Project, is interrupted in early stages, it yields no update. In later stages, it returns an arbitrage-free market state corresponding to a trade with a non-negative but possibly suboptimal profit. Thus, the loss bound is always maintained, even when \Project is interrupted.

\subsection{Fully-corrective Frank-Wolfe algorithm}
\label{sec:fw}

Recall that the FW algorithm reduces the problem of Bregman projection, i.e., a convex minimization over the set $\M$, into a sequence of linear optimization problems over the set $\Z$.
Our version, presented as \Algo{FW}, is based on the \emph{fully-corrective} variant of the Frank-Wolfe algorithm~\citep{Jaggi13}, also known as the \emph{simplicial decomposition method}~\citep{Bertsekas15:Convex}, which we overview next.

The FW algorithm solves problems of the form
\begin{equation}
\label{eq:FW}
   \min_{\vmu\in\M} F(\vmu)
\enspace,
\end{equation}
where $\M$ is a compact convex set (in our case a polytope)
and $F$ is a convex function. Over the course of iterations
$t=1,2,\dotsc$, the algorithm maintains
an active set $\Z_t$ of the vertices of the polytope $\M$ that have been discovered so far, and repeatedly:
\begin{enumerate}
\item
solves the minimization
over the convex hull of $\Z_{t-1}$ to obtain a new iterate
\[
 \vmu_t\coloneqq\argmin_{\vmu\in\conv(\Z_{t-1})} F(\vmu)
 \enspace,
\]
\item
finds a new descent vertex $\vz_t$ in the direction of the (negative) gradient of $F$,
\[
  \vz_t\coloneqq\argmin_{\vz\in\Z} \bigBracks{\nabla F(\vmu_t)\inprod\vz}
\enspace,
\]
\item
and adds $\vz_t$ to the set of active vertices, so $\Z_{t}=\Z_{t-1}\cup\set{\vz_t}$.
\end{enumerate}

Note that while the set $\Z$ of valid payoffs can be exponentially large, the set of active vertices $\Z_t$ grows by only one vertex per iteration (and is initialized with only a small number of vertices). Therefore,
Step~(1), which is a convex optimization problem of dimension $\card{\Z_t}$, can be solved efficiently by standard algorithms. We use
accelerated projected gradient~\citep{Nesterov07}.

Step (2), the linear optimization over the set $\Z$, is the computationally expensive step. As discussed in \Sec{fwmm:outline}, in our case it can be implemented by a call to an IP solver. In all of our experiments, the running time of Step~(2) substantially dominated the running time of Step~(1).

The convergence of the FW algorithm is analyzed via the FW gap, defined as
\[
  g(\vmu)\coloneqq \max_{\vz\in\Z} \bigBracks{\nabla F(\vmu)\inprod(\vmu-\vz)}
\enspace,
\]
which bounds the suboptimality of $\vmu$. Specifically, $g(\vmu)\ge F(\vmu)-F(\svmu)$, where $\svmu$ is a solution to \Eq{FW}. Thus, we can just monitor the gap $g(\vmu_t) = \nabla F(\vmu_t)\inprod(\vmu_t-\vz_t)$,
and return the iterate $\vmu_t$ when the gap becomes sufficiently
small. The gap converges to zero at the rate of $O(L\diam(\M)/t)$ where $L$ is the Lipschitz constant of $\nabla F$ under an arbitrary norm and $\diam(\M)$ is the diameter of $\M$ under the same norm~\citep{Jaggi13}.

To apply the FW algorithm to the problem of Bregman projection, we set its objective to the Bregman divergence: $F(\vmu)=\D{\vmu}{\vtheta}=R(\vmu) + C(\vtheta) - \vtheta\inprod\vmu$. One formal problem arises due to the fact that the function $R$ is not necessarily differentiable only subdifferentiable.
To overcome this, we assume existence of a differentiable extension~$\bR$. For LMSR, this is $\bR(\vmu)=\ind\set{\vmu\ge\vzero}+\sum_{i\in \I} \mu_{i}\ln \mu_{i}$, and similarly for the sum of \mbox{LMSRs}.
The key point is that $\bR$ coincides with $R$ over $\M$, so we can optimize the (differentiable) function $F(\vmu)=\bR(\vmu) + C(\vtheta) - \vtheta\inprod\vmu$. (More details in \App{growth}.)

Apart from differentiability, there are two additional challenges in applying the FW algorithm within \Mech{FWMM}.
First, we need to choose a stopping condition for the FW algorithm
that would yield a state update with a guaranteed profit, since such updates maintain the worst-case loss bound of the market maker.
Second,
even though we have achieved the differentiability of $F$ for our case of interest (the sum of \mbox{LMSRs}), the resulting derivative is unbounded, so the standard convergence analysis of FW does not apply. Fortunately, the growth of the derivative at the boundary is sufficiently controlled to obtain convergence of a
modified version of FW, which is what we use in \Algo{FW}.
(The precise statement of the controlled growth condition is in \App{growth}.)

\begin{algorithm}[t]
\caption{$\Project$. Bregman Projection via Adaptive Fully-Corrective Frank-Wolfe.}
\label{algo:FW}
\begin{tabbing}
\textbf{Input:~~~}%
  cost function $C$, state $\vtheta$, partial outcome $\sigma$,\\
\hphantom{\textbf{Input:~~~}}%
  IP constraints specified by $\A$, $\vb$,\\
\hphantom{\textbf{Input:~~~}}%
  approx.\ ratio $\alpha\in(0,1)$, initial contraction $\eps_0\in(0,1)$, convergence threshold $\eps_D>0$\\[4pt]
\textbf{Output:~~~}%
  extended partial outcome $\hsigma\supseteq\sigma$\\
\hphantom{\textbf{Output:~~~}}%
  state $\hvtheta$, whose price vector is an approx.\ Bregman projection of $\vtheta$ on $\M_\hsigma$ in the sense\\
\hphantom{\textbf{Output:~~~}}%
  that one of the following holds:\\
\hphantom{\textbf{Output:~~~\quad}}%
  1. $\vp_\hsigma(\hvtheta)\in\M_\hsigma$ \
     and moving from $\vtheta$ to $\hvtheta$ guarantees the profit of $\alpha\D[\hsigma]{\svmu}{\vtheta}$\\
\hphantom{\textbf{Output:~~~\quad}}%
  2. $\hvtheta=\vtheta$ and $\D[\hsigma]{\svmu}{\vtheta}\le\eps_D$\\
\hphantom{\textbf{Output:~~~\quad}}%
  3. algorithm was interrupted; moving from $\vtheta$ to $\hvtheta$ guarantees a non-negative profit\\
\hphantom{\textbf{Output:~~~}}%
  where $\svmu=\argmin_{\vmu\in\M_\hsigma}\D[\hsigma]{\vmu}{\vtheta}
  \vphantom{\hvtheta}$\\[4pt]
  Initialize the interior point, active vertex set, and extend the partial outcome:\\
\hphantom{---}\=
    $(\vu,\Z_0,\hsigma)\gets\InitFW(\sigma,\A,\vb)$\\[4pt]
  Define the objective function:\\
\>
    $F(\vmu)\coloneqq \bR_\hsigma(\vmu) - \vtheta\inprod\vmu + C_\hsigma(\vtheta)$\\[4pt]
  For $t=1,2,\dotsc$\\[4pt]
\hphantom{---}\=
    \emph{perform a FW iteration on the contracted polytope:}\\
\>\hphantom{---}\=
        let $\strut\Z'=(1-\eps_{t-1})\Z_{t-1} +\eps_{t-1}\vu\;$ denote the contracted active set\\
\>\>
    $\vmu_t\gets\argmin_{\vmu\in\conv(\Z')} F(\vmu)$\\
\>\>
    $\vtheta_t \gets \nabla\bR_\hsigma(\vmu_t)$\\
\>\>
    call IP solver to find the descent vertex (note that $\nabla F(\vmu_t)=\vtheta_t-\vtheta$):\\
\>\>\hphantom{---}\=
        $\vz_t\gets\argmin_{\vz\in\Z_{\hsigma}} (\vtheta_t-\vtheta)\inprod\vz$\\
\>\>
    $\Z_t=\Z_{t-1}\cup\set{\vz_t}$\\[4pt]
\>
    compute the FW gap $g(\vmu_t)=(\vtheta_t-\vtheta)\inprod(\vmu_t-\vz_t)$\\
\>
    update the best-iterate-so-far
    $t^\star\gets\argmax_{\tau\le t} \bigBracks{F(\vmu_\tau)-g(\vmu_\tau)}$\\[4pt]
\>
    \emph{check stopping conditions:}\\
\>\>
        if $g(\vmu_t)\le(1-\alpha) F(\vmu_t)$,\;\;
            or\;
           $F(\vmu_t)\le\eps_D$,\;\;
            or\;
            termination requested\\[2pt]
\>\>\>
            return $\hsigma$ and $\hvtheta=
            \begin{cases}
            \vtheta_{t^*}&\text{if $g(\vmu_{t^*})\le F(\vmu_{t^*})$}
            \\
            \vtheta&\text{otherwise}
            \end{cases}$\\[2pt]
\>
    \emph{adapt contraction if necessary:}\\
\>\>
        let $g_\vu=(\vtheta_t-\vtheta)\inprod(\vmu_t-\vu)$\\
\>\>
        if $g_\vu < 0$ and $g(\vmu_t)/(-4g_\vu) <\eps_{t-1}$\\
\>\>\>
            $\eps_t\gets\min\bigSet{ g(\vmu_t)/(-4g_\vu),\,\eps_{t-1}/2 }$\\
\>\>
        else\\
\>\>\>
            $\eps_t\gets\eps_{t-1}$\\[-16pt]
\end{tabbing}
\end{algorithm}

The modified version of FW, due to~\citet{KrishnanLaSo15}, performs FW iterations over a contracted version of the polytope $\M$, or, more precisely, over a contracted version of $\M_{\hsigma}$, which reflects already settled securities. The contracted polytope is defined as $\M'\coloneqq(1-\eps)\M_{\hsigma} + \eps\vu$, where $\vu\in\M_{\hsigma}$ is a coherent price vector
whose coordinates are neither 0 nor 1, except for those already settled by $\hsigma$. In other words, $\M'$ is a version of $\M_{\hsigma}$ shrunk towards the point $\vu$, which we call an \emph{interior point}. Since coordinates of $\vu$ are bounded away from 0 and 1,
the vertices of the contracted polytope~$\M'$ have their coordinates also bounded away from 0 and 1 (except for $\I_\hsigma$). The controlled growth property
then gives a bound on the Lipschitz constant of the gradient and guarantees convergence for any fixed $\eps$, for the problem of projecting onto $\M'$. To obtain the convergence to the projection onto $\M_{\hsigma}$, we adaptively decrease $\eps$ according to the rule of~\citet{KrishnanLaSo15}. Their analysis shows that this adaptive version of FW drives the duality gap $g(\vmu_t)$ to zero and thus indeed solves the non-contracted problem. Two missing pieces that we describe in the remainder of this section are the stopping condition and the construction of the interior point $\vu$.

\subsection{Stopping condition for the FW algorithm}
\label{sec:stopping}

The stopping condition needs to ensure that moving the market from a state $\vtheta$ to $\hvtheta$ constitutes a trade with a non-negative profit.
We start with a lower bound on the guaranteed profit of any iterate of the FW algorithm,
and then use it to derive the stopping condition. We omit the conditioning on $\hsigma$
from the exposition here.
%

\begin{proposition}
   Consider a purchase that moves the market from a state $\vtheta$ to a new state $\hvtheta=\nabla\bR(\hvmu)$. The resulting profit
   is guaranteed to be at least
  $\D{\hvmu}{\vtheta} - g(\hvmu)$.
\label{prop:gap}
\end{proposition}


Thus, it is ``safe'' to move the market to $\hvtheta$ whenever $\D{\hvmu}{\vtheta}\ge g(\hvmu)$ (for proof see \App{proof:gap}). To maximize the profit guarantee, we should return the iterate that maximizes the difference
$\D{\hvmu}{\vtheta}-g(\hvmu)$, which is what we do in \Algo{FW}.

Apart from a forced interruption (e.g., because of the arrival of a new trade or exceeding of the time limit), the stopping conditions of \Algo{FW} concern two separate cases. First,
recall that the algorithm is minimizing $F(\vmu)=\D{\vmu}{\vtheta}$
via a sequence of iterates $\vmu_t\in\M$
that satisfy $\D{\vmu_t}{\vtheta}\to\D{\svmu}{\vtheta}$
and $g(\vmu_t)\to 0$ as $t\to\infty$.
Therefore, if prices $\vp(\vtheta)$ are incoherent, i.e., $\D{\svmu}{\vtheta}>0$, eventually
we will have $g(\vmu_t)<\D{\vmu_t}{\vtheta}$. In fact, we can guarantee something stronger. Namely, given a fixed $\alpha\in(0,1)$, we will reach an iteration when
\[
  g(\vmu_t)\le(1-\alpha)\D{\vmu_t}{\vtheta}
\enspace.
\]
At this point, our profit guarantee is at least
\[
  \D{\vmu_t}{\vtheta}-g(\vmu_t)\ge \alpha\D{\vmu_t}{\vtheta}\ge \alpha\D{\svmu}{\vtheta}
\]
thanks to the optimality of $\svmu$. This means that we are extracting at least an $\alpha$-fraction of the available arbitrager profits; this covers the first stopping condition and the first output case of \Algo{FW}.
On the other hand, if the prices $\vp(\vtheta)$ are coherent or close-to-coherent, then $\D{\vmu_t}{\vtheta}$ will eventually drop below our convergence threshold $\eps_D$, which we can set arbitrarily small. Since
$\D{\svmu}{\vtheta}\le\D{\vmu_t}{\vtheta}$, this covers the second stopping condition and the second output case of \Algo{FW}. The final case follows directly from \Prop{gap}.


\begin{algorithm}[t]
\caption{$\InitFW$. Initialization for \Project.}
\label{algo:InitFW}
\begin{tabbing}
\textbf{Input:\quad}%
  partial outcome $\sigma$, IP constraints specified by $\A$, $\vb$\\[4pt]
\textbf{Output:\quad}%
  extended partial outcome $\hsigma\supseteq\sigma$\\
\hphantom{\textbf{Output:\quad}}%
  point $\vu\in\M_\hsigma$ such that $u_i\in(0,1)$ for $i\not\in\I_{\hsigma}$\\
\hphantom{\textbf{Output:\quad}}%
  non-empty set $\Z_0$ of vertices of $\M_\hsigma$\\[4pt]
Initialize $\Z_0\gets\emptyset$, $\hsigma\gets\sigma$, $\C\gets\emptyset$\\[4pt]
For each $i\in\I\wo\I_\sigma$ and each $b\in\set{0,1}$\\
\hphantom{---}\=
    if $(i,b)\not\in\C$\\
\>\hphantom{---}\=
        call IP solver to find $\hvz=\argmax_{\vz\in\Z_{\sigma}} (2b-1)z_i$\\
\>\>
        if $\hz_i=b$\\
\>\>\hphantom{---}\=
            $\Z_0\gets\Z_0\cup\set{\hvz}$\\
\>\>\>
            $\C\gets\C\cup\set{(j,\hz_j):\:j\in\I}$\\
\>\>
        else\\
\>\>\>
            $\hsigma\gets\hsigma\cup\set{(i,1-b)}$\\[4pt]
If $\Z_0=\emptyset$\\
\>
   $\Z_0\gets\set{\text{the unique point compatible with $\hsigma$}}$\\
Return $\hsigma$, $\Z_0$, and $\vu=\frac{1}{\card{\Z_0}}\sum_{\vz\in\Z_0}\vz$\\[-16pt]
\end{tabbing}
\end{algorithm}

\subsection{Finding the interior point}
\label{sec:contraction-point}

The goal here is to find a point $\vu\in\M$ where coordinates corresponding to unsettled securities
are strictly between 0 and 1. In the process, we also obtain the initial set of active vertices and an extended partial outcome $\hsigma$. To construct $\vu$, \Algo{InitFW} iterates through coordinates $i$ that have not been settled in the provided partial outcome $\sigma$, and calls the IP solver to find a valid vector $\hvz$ that is consistent with $\sigma$, but also has the $i$-th coordinate equal to $b=0$ or $b=1$. If the IP solver fails to find such $\hvz$ for either value $b$, it means that the $i$-th coordinate can be settled to $1-b$. Otherwise the found $\hvz$ is added to the set of active vertices. This guarantees that each coordinate $i$ is either present in~$\hsigma$, or the active set contains some valid vertices with both the value $0$ and $1$ at the $i$-th coordinate.
Therefore, the average of the active vertices satisfies the requirement for~$\vu$.
If the active set is empty, it means that all of the securities have been settled and the unique valid vector consistent with $\hsigma$ satisfies the requirement.

%% file: dataset.tex

\section{Experiments}
\label{sec:exp}

\subsection{Data description}

Our data consists of bets made in \Predictalot, a combinatorial
prediction market run by Yahoo!\ in 2010 for the NCAA Men's Division I
Basketball Tournament, commonly known as \emph{March
  Madness}.\footnote{ Securities in the \Predictalot market were
  priced using the Monte Carlo method with importance sampling against
  a dynamic proposal distribution. One of the larger issues, which we
  do not expect with the optimization methods presented in this work,
  was substantial price volatility as the tournament progressed, due
  to an increasing mismatch between the market belief and the proposal
  distribution. In order to avoid trivial arbitrage, independent
  samples were drawn to form the prices quoted to the traders, and the
  actual prices imposed on trade executions. As a result, some trades
  transacted at prices significantly different than quoted. [D.
  Pennock, personal communication, Feb.\ 22, 2016]}
The tournament lasted from March 18th to April 5th, 2010. It consisted
of 64 teams playing a single-elimination tournament over 6 rounds. In
each round, half of the remaining teams were eliminated. Traders were
allowed to buy securities at any point in time throughout the
tournament; the first bets were placed four days prior to the
tournament start
and the last bets were placed towards the end of the final
match. Many bets referred to groupings of teams, known as conferences,
brackets or seeds (e.g., there are sixteen seed levels and four teams
to each seed).

There were 93\,036 bets placed
altogether on many different securities in \Predictalot.
Our experiments
focus on a large subset of these, which we briefly describe here. The largest group
of bets (56\%) can be expressed as bundles over atomic tournament variables (winners of individual games,
and the number of wins of individual teams). These include bets such as ``\Duke wins exactly 3 games'', ``\Cornell exits in round 2 or later'', ``a team from the \emph{Big Ten} conference wins the championship''. In addition to these bets, we also supported combinatorial
bets for comparisons of the number of wins of single teams, e.g., ``\Duke wins more games
than \Cornell'', and comparisons of the number of wins by teams from different
conferences, e.g., ``teams from \emph{Big Ten} win more games than teams from \emph{Big East}''. These were implemented as comparison variables derived from pairs of atoms,
and pairs of sums, respectively.
The two comparison types encompass 12\% of original bets.

Our resulting dataset contains 63\,689 bets, constituting
68\% of all bets in the original market. Combinatorial bets (comparisons) make up
17\% of our final dataset. The three largest groups of bets we did not
include were:
``team $t_1$ wins more games than $t_2$, and $t_3$ wins
more games than $t_4$'' (6\%);
``the number of upsets in round $r$ will be less than/equal
to/greater than $c$'' (3\%);
and ``the sum of seeds in round $r$ will be less than/equal to/greater
than $c$'' (3\%).

\paragraph{Price initialization}
Our dataset contains realized trades, but we have no other price data from the run of the market.
In particular, the initial \Predictalot prices were not available, so we used the following
scheme to initialize atomic tournament variables $X_t$ (the number of wins of team $t$) and
$G_{r,t}$ (the outcome of a game). We considered bets within the 6 hour time window starting at 27 hours and ending at 21 hours before the first match of the tournament. Let $\mu'$ denote the price at which securities were sold in this window (we use last such price if multiple exist).
To initialize the game variables $G_{r,t}$, we use the prices of bets on the champion of the tournament (i.e., $X_t=k$):
\[
  \mu\set{ G_{r,t}=t }
  = \frac{\mu'\set{ X_t=k }}{\sum_{t'\in T}\mu'\set{ X_{t'}=k }}
\enspace,
\]
where $T$ is the set of all teams that can reach the game $G_{r,t}$; if the denominator equals zero, we initialize prices $\mu\set{ G_{r,t} = t'}$ across $t'\in T$ to a uniform distribution.
To initialize securities $X_t=x$, we proceeded as follows. If $\mu'\set{X_t=x}$ is present, we use that as the initialization price, otherwise we use the difference between $\mu'\set{G_{x,t}=t}$ and $\mu'\set{G_{x+1,t}=t}$, where we replace one or both of these terms by our already calculated prices according to $\mu$ whenever the $\mu'$ prices are not present. The resulting prices
are then normalized to sum to one for each $X_t$. The team and game prices are then projected on the polytope described by LCMM constraints to obtain market initialization.

\paragraph{Settling outcomes}
Similar to initialization prices, the times when the individual games were settled were not available, so we
%
%
handcrafted a dataset consisting of all game start times\footnote{Source: espn.com, e.g., \url{http://scores.espn.go.com/ncb/boxscore?gameId=300950150}} (to the best of our knowledge, end times are not listed anywhere) and settled each game 100 minutes after the game start. The choice of 100 minutes is conservative, based on the anecdotal observation that the shortest NCAA games last about 120 minutes, including the time for commercials and timeouts.
%
%



%% file: eval.tex

\subsection{Evaluation}
\label{sec:eval}

We compare three market treatments: independent markets (IND), the
linearly contrained market maker (LCMM), and a market maker with both
linear constraints and Bregman projections for arbitrage removal
(FWMM). Each market maker builds upon and extends the previous one.
Recall that in IND, we use LMSR
to price the
securities associated with each random variable, but prices for separate
variables vary independently, even if the underlying events are
related. LCMM enforces price relationships across random variables
using linear constraints, and FWMM adds projection steps onto the
marginal polytope.
The market makers were implemented in Java, using Gurobi Optimizer
5.5\footnote{%
  \url{www.gurobi.com}} to solve the integer programs in the
FW algorithm. We refer to our implementation as the (market)
engine.

We evaluate the three market makers by a counterfactual replay of the trades
placed in \Predictalot. All the market makers depend on the liquidity parameter $b$~(see Eq.~\ref{eq:lmsr-cost}). Rather than optimizing $b$,
%
we used a fixed liquidity of 150 and
varied each trader's budget. (The effect
is equivalent, as increasing the budget increases price responsiveness
to the trade orders.) Each trade order is viewed as a new agent, so
the budget is constant for each trade. We used budget levels 0.1, 1,
10, 100, and 1000.

For each trade, the \Predictalot dataset contains the
number of shares purchased and the total cost paid. By
taking the average price per share $\bar{p}$, we obtain a \emph{lower
  bound} on the trader's probability estimate when the trade was
placed. From this we create a limit order for our market engine by
drawing a limit price uniformly from $[\bar{p}, 1]$, and providing the
constant budget level mentioned previously. A limit order states that
the trader wishes to purchase shares until either the market price
reaches the limit price, or the budget is exhausted, whichever occurs
first.
Any sell orders with average price $\bar{p}$ were transformed into
buy orders of the complementary bundle, at price $1-\bar{p}$, and then
converted into limit orders.
By using three different seeds for the randomization, we generated
three input files for the market engine. All market makers were run on
all three input files. As the results were highly consistent across
the randomization seeds, we found three replicates to be sufficient.

To summarize, we ran the three different market makers (IND, LCMM,
FWMM) at five budget levels (0.1, 1, 10, 100, 1000) over the three
randomly generated input files. During a market run, the engine records
summaries of security prices and prices of all purchased bundles.
These summaries
are generated at regular intervals, including
every hour and every 100 trades.
%
%
We use the
\emph{log likelihood} to assess the accuracy of the security prices, viewed as
probability forecasts, at a given point in time. Let $\vmu$ be the price vector. We consider log likelihoods associated with two different kinds of events. First is the log likelihood assigned to the final realized value $x^\star$ of a variable $X$, which equals
$
  \log\mu\set{X=x^\star}
$.
%
Second is the log likelihood corresponding to the bundle of the form $X\in\Eset$, viewed as a binary variable (the event occurs or not), which is defined as
\[
  1\set{x^\star \in \Eset} \log \mu\set{X \in \Eset}
  + 1\set{x^\star\not\in \Eset} \log \mu\set{X \not\in\Eset}
    \enspace.
\]
A larger log likelihood indicates a better forecast. We report the average log likelihood over all variables, and the average log likelihood over all purchased bundles. The former can be viewed as an average accuracy of the market, the latter is weighted towards the part of the market that sees more trading.

\begin{figure}
\centering
  \includegraphics[scale=.388]{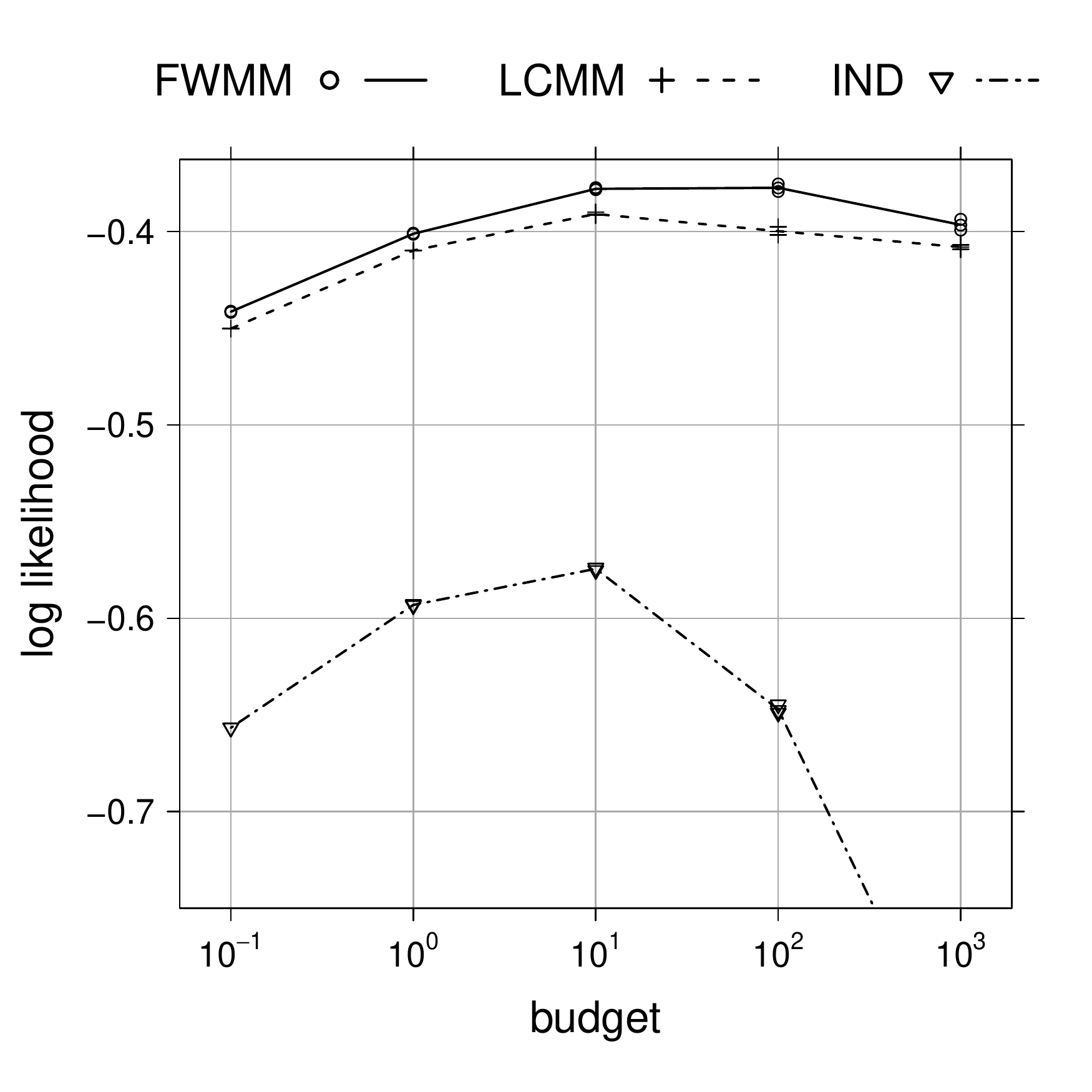}
  \includegraphics[scale=.388]{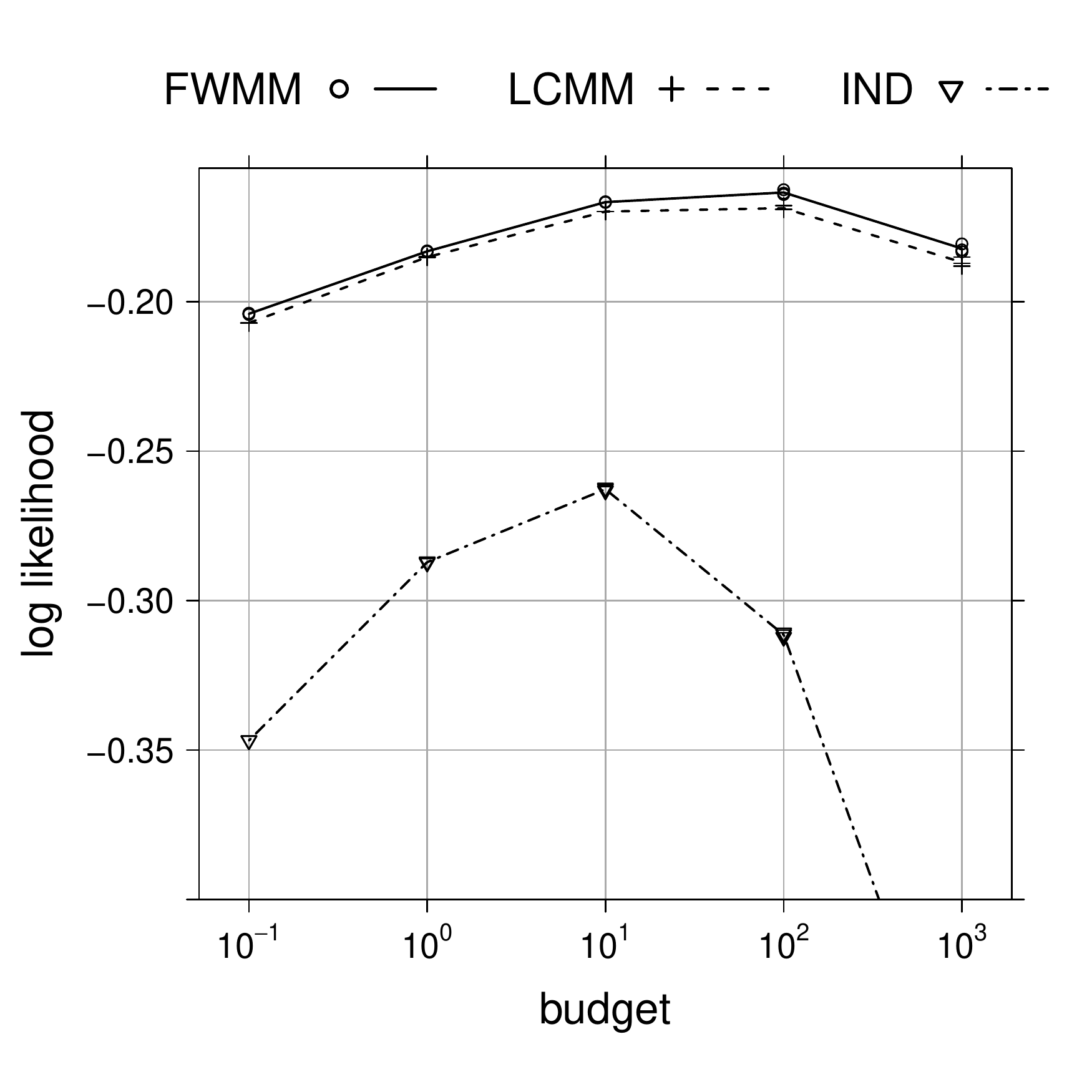}
\figsqueezeR%
\caption{Market maker accuracy, varying the budget level. \emph{Left:} average over variables.
\emph{Right:} average over bundles. Average taken over all
hourly logs and all random seeds.}
\label{fig:liquidity}
\figsqueeze%
\end{figure}

\paragraph{Effect of liquidity} We first examine the effect of varying the
budget level (equivalently, liquidity) on the overall performance of
the three market makers. \Fig{liquidity} provides the
average prediction accuracy of the three market
makers over variables and bundles, where the average is taken over all hourly
summaries. The plots show the expected
trends: when budget is too low, traders cannot incorporate their
information into the market, while when budget is too high, prices are
too sensitive to individual trades. The optimal budget setting is 10
for IND and LCMM, and 100 for FWMM. However, both LCMM and FWMM are
far less sensitive to the budget level than IND, because information
propagation (via constraints) can correct wrong bets.


The improvement of FWMM over LCMM for variables ranges from
$2.1\%$ to $5.6\%$, with a median of $3.3\%$ over all budget levels and
random seeds. For bundles, the improvement ranges from
$0.9\%$ to $3.2\%$, with a median of $2.2\%$. For the time period
covering the first 16 games, LCMM and FWMM are very similar (see the
next section), bringing their average performance closer together;
excluding these games, the median improvement increases from $3.3\%$
to $12.4\%$ for securities, and from $2.2\%$ to $5.6\%$ for bundles.
Because accuracy here is averaged over all hourly summaries, it is
implictly weighed by duration, which is hard to interpret. To obtain a
more fine-grained view, we next consider the evolution of market
accuracy over time.

%

\paragraph{Accuracy over time}

\begin{figure}
\centering
  \includegraphics[scale=.388]{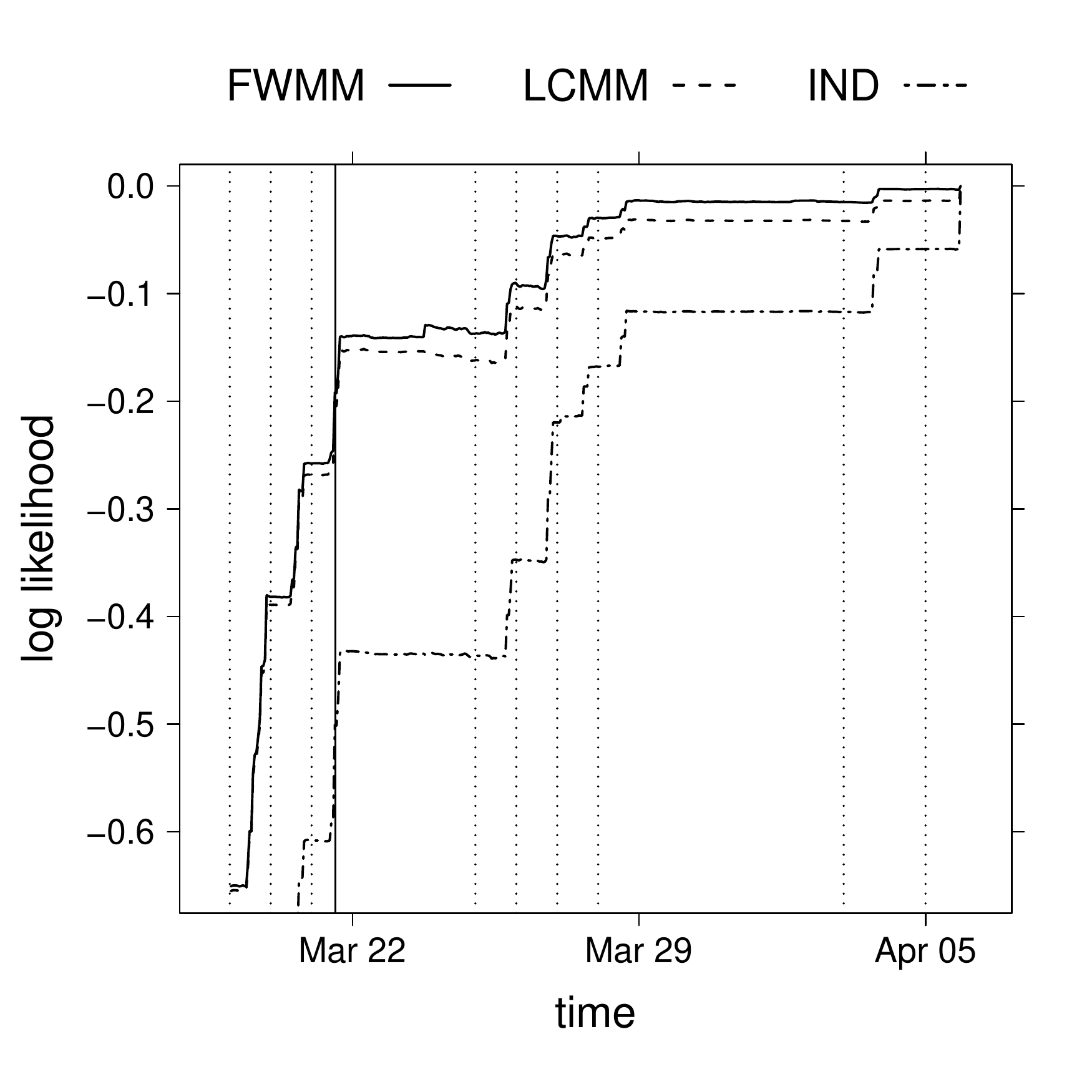}
  \includegraphics[scale=.388]{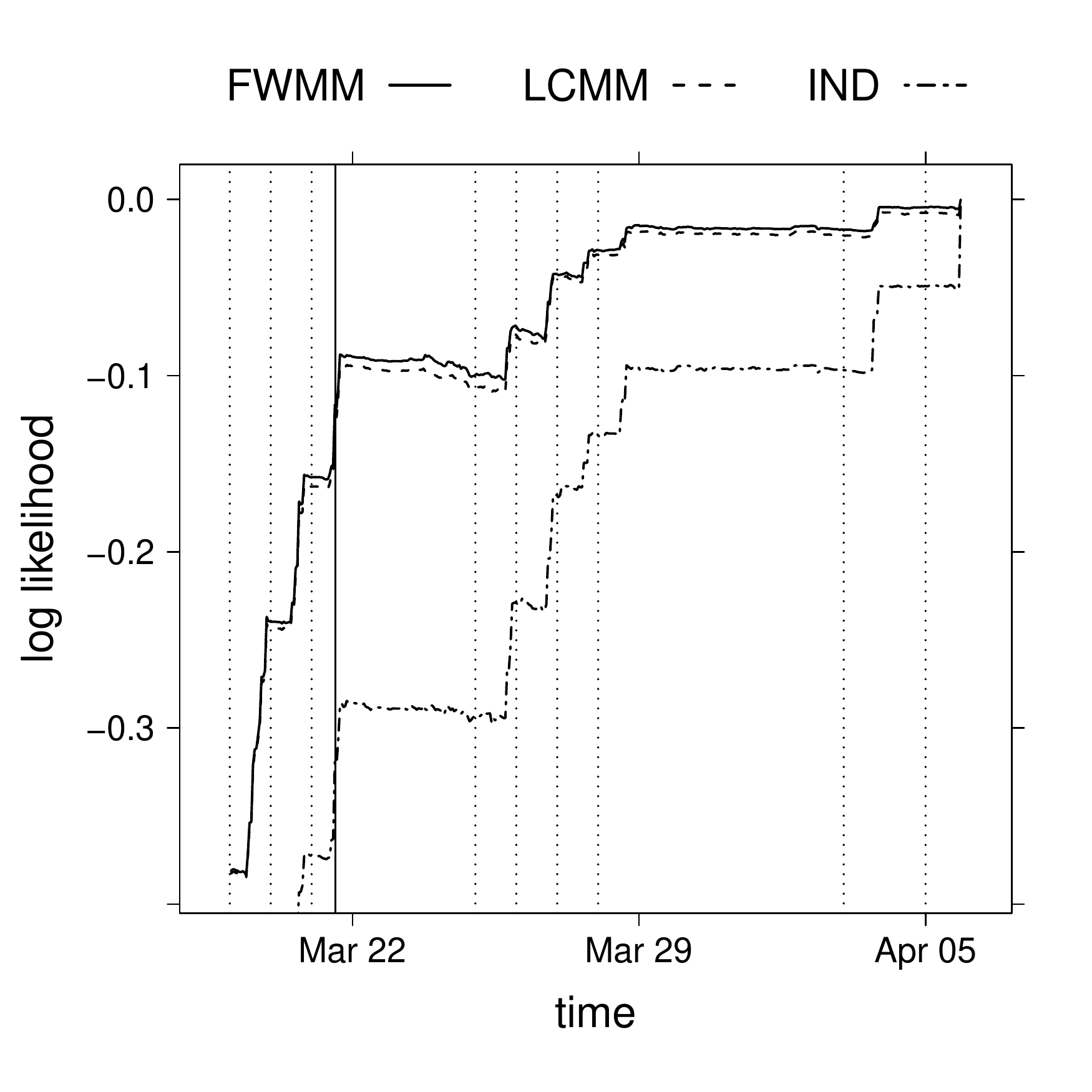}
\figsqueezeR%
\caption{Market maker accuracy over time, at budget level 10.
  \emph{Left:} average over variables. \emph{Right}: average over
  bundles. Average taken over all variables or bundles defined at that
  time, over all random seeds. The dotted vertical lines are at 00:00 on
  days when games are played. The solid vertical line indicates the
  start of projections in FWMM.}
\label{fig:time}
\figsqueeze%
\end{figure}

\Fig{time} plots the prediction accuracy of the three
markets as time progresses. We set a time limit of 30 minutes
for Bregman projection. The first time it successfully completes
is only
at time stamp \mbox{`2010-03-21 13:58:50'}, after
45 games are already settled.
We therefore begin the plot at time stamp \mbox{`2010-03-19
  00:00:00'}, corresponding to 16 settled games,
as there is very little difference between LCMM and FWMM
before that point. The reason FWMM still exceeds LCMM on occasion
before the first projection is due to the extension of the partial outcome
afforded by the IP, as explained in \Sec{fwmm}.

Each point of the time series represents an average over all
variables or bundles defined at that time, including those whose
outcomes have been settled. This explains the upwards trends of the
plots, culminating at accuracy 0 (a perfect score). The trend is not
entirely monotonic, as we see from the bundle log likelihood in the
stretch after March 22. The dotted vertical lines indicate the beginning
of days on which games are played. On such days, we see that accuracy
is initially stable, then sharply increases as the games take course
and their outcomes are settled.

In \Fig{time}, we see that once Bregman projections successfully complete, the
improvement of FWMM over LCMM becomes sustained. The accuracy
improvements from this point onwards range from $0\%$ to $80\%$ for
variables, with a median of $38\%$ over all hourly summaries.
The improvements range from $0\%$ to $44\%$ for bundles, with a median of
$9\%$.


%

%% file: conclusion.tex
\section{Discussion and Conclusion}
\label{sec:conclusion}

In our experiments, FWMM outperformed LCMM once the outcome space was sufficiently reduced, via settled securities, to allow computing of Bregman projections within 30 minutes on a standard workstation. This time limit yielded a manageable experimental turnaround, with about 5 hours to execute the trades that originally spanned 22 days. In practice, a market designer can allow longer computation and use more powerful hardware, and expect improvements for larger problem sizes.

Several approaches could further speed up our framework. For instance, FW can be used to construct separating hyperplanes to tighten the outer LCMM approximation, and thereby contribute to arbitrage removal even when there is no time to compute the projection.
Also, instead of solving IPs to optimality in each iteration, it may be possible to interleave IP with local search to obtain additional descent vertices.
Since IP is by far the most time-consuming part of FW, this could yield substantial speedups.

%% file: appendix.tex
\appendix
\section*{APPENDIX}

\setcounter{section}{0}

\section{Proof of Proposition 2.4}
\label{app:proof:arb:proj}

We first calculate the largest possible guaranteed profit:
\begin{align}
\notag
\adjustlimits\sup_{\vdelta\in\R^\I}\min_{\vz\in\Z}
   \bigBracks{ \vdelta\inprod\vz-C(\vtheta+\vdelta)+C(\vtheta) }
&=
\notag
\adjustlimits\sup_{\vtheta'\in\R^\I}\min_{\vmu\in\M}
   \bigBracks{ (\vtheta'-\vtheta)\inprod\vmu-C(\vtheta')+C(\vtheta) }
\\
&=
\label{eq:arb:1}
\adjustlimits\min_{\vmu\in\M}\sup_{\vtheta'\in\R^\I}
   \bigBracks{ (\vtheta'-\vtheta)\inprod\vmu-C(\vtheta')+C(\vtheta) }
\\
&=
\label{eq:arb:2}
\min_{\vmu\in\M}
   \bigBracks{ R(\vmu)-\vtheta\inprod\vmu + C(\vtheta) }
\\
&=
\label{eq:arb:3}
\min_{\vmu\in\M} \D{\vmu}{\vtheta} = \D{\svmu}{\vtheta}
\enspace,
\end{align}
where \Eq{arb:1} follows by Sion's minimax theorem and Eqs.~\eqref{eq:arb:2}
and~\eqref{eq:arb:3} from definitions of the convex conjugate and Bregman
divergence, respectively. This shows that from the state $\vtheta$ the guaranteed profit
is at most $\D{\svmu}{\vtheta}.$

Recall that $\svdelta$ is any trade that moves the market to a state $\svtheta$ such that
$\vp(\svtheta)=\svmu$.
We
next show that $\svdelta$ is an optimal trade, i.e., that this trade
gives a profit that is at least $\D{\svmu}{\vtheta}.$
Let $F(\vmu)\coloneqq\D{\vmu}{\vtheta}$.
Since $\svmu$ optimizes $F$ on $\M$, by the first order optimality, we have
for any $\vu\in\partial F(\svmu)$ and $\vz\in\Z$ that
$\vu\inprod(\vz-\svmu)\ge 0$. Since $\vp(\svtheta)=\svmu$, the conjugacy implies
that $\svtheta\in\partial R(\svmu)$ and thus $(\svtheta-\vtheta)\in\partial F(\svmu)$,
so the first order optimality yields
\[
  0\le(\svtheta-\vtheta)\inprod(\vz-\svmu)
\enspace,
\]
which rearranges to
\begin{equation}
\label{eq:arb:4}
 (\svtheta-\vtheta)\inprod\vz\ge(\svtheta-\vtheta)\inprod\svmu
\enspace.
\end{equation}
The profit from the trade $\svdelta$ given any outcome $\omega$ is therefore at least
\begin{align*}
 (\svtheta-\vtheta)\inprod\vphi(\omega)-C(\svtheta)+C(\vtheta)
& \ge
 (\svtheta-\vtheta)\inprod\svmu-C(\svtheta)+C(\vtheta)
\\
& =
  R(\svmu)-\vtheta\inprod\svmu+C(\vtheta)
\\
& =
  \D{\svmu}{\vtheta}
\enspace,
\end{align*}
where the first line follows by substituting $\vphi(\omega)$ for $\vz$ in \Eq{arb:4}, the second line follows from the conjugacy of $R$ and $C$, and the third line
from the definition of $D$,
completing the proof.
\qed

\section{Bounded loss property under gradual revelation of outcome}
\label{app:partial}

We show that the bound on the worst-case loss of the cost $C$ is maintained if we update the cost function using a sequence of partial outcomes, gradually revealing the final outcome $\omega$. We begin with the worst-case bound on the loss under cost $C$:

\begin{proposition}
\label{prop:wc:C}
If the initial market state is $\vtheta_0$ then the worst-case loss of a market-maker using $C$ is $\max_{\omega\in\Omega} \D{\vphi(\omega)}{\vtheta_0}$.
\end{proposition}
\begin{proof}
Let $\vtheta$ denote the final state before the outcome $\omega$ is revealed. Then the market maker has collected $C(\vtheta)-C(\vtheta_0)$ as the revenue for the sold shares, and needs to pay out $(\vtheta-\vtheta_0)\inprod\vphi(\omega)$ as a payoff to the traders. The worst-case loss is therefore
\begin{align}
\notag
&
\adjustlimits\max_{\omega\in\Omega}\sup_{\vtheta}
\BigBracks{
  (\vtheta-\vtheta_0)\inprod\vphi(\omega) - \bigParens{C(\vtheta)-C(\vtheta_0)}
}
\\
\notag
&\quad{}
=
\adjustlimits\max_{\omega\in\Omega}\sup_{\vtheta}
\BigBracks{
  \bigParens{\vtheta\inprod\vphi(\omega)-C(\vtheta)}
  -\vtheta_0\inprod\vphi(\omega) +C(\vtheta_0)
}
\\
\notag
&\quad{}
=
\max_{\omega\in\Omega}
\BigBracks{
  R\bigParens{\vphi(\omega)}
  -\vtheta_0\inprod\vphi(\omega) +C(\vtheta_0)
}
\\
\tag*{\qed}
&\quad{}
=
\max_{\omega\in\Omega} \D{\vphi(\omega)}{\vtheta_0}
\enspace.
\end{align}
\renewcommand{\qed}{}%
\end{proof}

Now, we will analyze the case with partial outcomes.
We assume that the initial partial outcome $\sigma_0=\emptyset$, and that the market goes through a sequence of partial outcomes $\sigma_0\subseteq\sigma_1\subseteq\dotsb\subseteq\sigma_T$ until finally an outcome $\omega$ is revealed, consistent with $\sigma_T$. After the revelation of each $\sigma_t$, the market-maker switches to the cost function $C_{\sigma_t}$. The initial market state is denoted $\vtheta_0$ and the market state in which the market switches to $C_{\sigma_t}$ is denoted $\vtheta_t$.

\begin{proposition}
If the initial market state is $\vtheta_0$ then, regardless of the sequence of partial outcomes $\sigma_1,\dotsc,\sigma_T$, the worst-case loss of the market-maker using the sequence of costs $C_{\sigma_t}$ is $\max_{\omega\in\Omega} \D{\vphi(\omega)}{\vtheta_0}$, i.e., the same as that of the market-maker using $C$ without incorporating partial outcomes.
\end{proposition}
\begin{proof}
Recall that the market state at the time of switch from $C_{\sigma_{t-1}}$ to $C_{\sigma_{t}}$ is $\vtheta_t$. We first show that the value of the cost at the time of switch decreases:
\begin{equation}
\label{eq:Csigma:t}
 C_{\sigma_t}(\vtheta_t)
=
  \sup_{\vmu\in V_{\sigma_t}} \Bracks{\vtheta\inprod\vmu-R(\vmu)}
\le
  \sup_{\vmu\in V_{\sigma_{t-1}}} \Bracks{\vtheta\inprod\vmu-R(\vmu)}
=
 C_{\sigma_{t-1}}(\vtheta_t)
\end{equation}
where the middle inequality follows because $V_{\sigma_t}\subseteq V_{\sigma_{t-1}}$.
We are now ready to prove the bound on the worst-case loss. Let $\Omega(\sigma_T)$ denote the set of outcomes compatible with $\sigma_T$, and recall that $\sigma_0=\emptyset$, so $C_{\sigma_0}\equiv C$. Recall that $\vtheta_t$ for $t=1,\dotsc,T$ are the states of the market when the cost becomes $C_{\sigma_t}$. Finally, let $\vtheta_{T+1}$ denote the final state. Then the worst-case loss of the market maker can be bounded as follows
\begin{align}
\notag
&
\max_{\sigma_1\subseteq\sigma_2\subseteq\dotsc\subseteq\sigma_T\vphantom{\Omega}}
\adjustlimits\max_{\omega\in\Omega(\sigma_T)}\sup_{\vtheta_1,\dotsc,\vtheta_{T+1}}
\Bracks{
  (\vtheta_{T+1}-\vtheta_0)\inprod\vphi(\omega) - \sum_{t=0}^T\bigParens{C_{\sigma_t}(\vtheta_{t+1})-C_{\sigma_t}(\vtheta_t)}
}
\\
\notag
&\quad{}=
\max_{\sigma_1\subseteq\sigma_2\subseteq\dotsc\subseteq\sigma_T\vphantom{\Omega}}
\adjustlimits\max_{\omega\in\Omega(\sigma_T)}\sup_{\vtheta_1,\dotsc,\vtheta_{T+1}}
\biggl[
  (\vtheta_{T+1}-\vtheta_0)\inprod\vphi(\omega)
  - \bigParens{C_{\sigma_T}(\vtheta_{T+1})-C_{\sigma_0}(\vtheta_0)}
\biggr.
\\
\label{eq:rev:1}
&\hspace{2.25in}\biggl.{}
  -
  \sum_{t=1}^T\bigParens{C_{\sigma_{t-1}}(\vtheta_{t})-C_{\sigma_t}(\vtheta_t)}
\biggr]
\\
\label{eq:rev:2}
&\quad{}\le
\max_{\sigma_T\vphantom{\Omega}}
\adjustlimits\max_{\omega\in\Omega(\sigma_T)}\sup_{\vtheta_{T+1}}
\BigBracks{
  (\vtheta_{T+1}-\vtheta_0)\inprod\vphi(\omega)
  -C_{\sigma_T}(\vtheta_{T+1})+C_{\sigma_0}(\vtheta_0)
}
\\
\label{eq:rev:4}
&\quad{}=
\max_{\sigma_T\vphantom{\Omega}}\max_{\omega\in\Omega(\sigma_T)}
\BigBracks{
  R\bigParens{\vphi(\omega)}
  -\vtheta_0\inprod\vphi(\omega)
  +C(\vtheta_0)
}
\\
\label{eq:rev:5}
&\quad{}=
\max_{\omega\in\Omega} \D{\vphi(\omega)}{\vtheta_0}
\enspace.
\end{align}
\Eq{rev:1} follows by rearranging the terms. \Eq{rev:2} follows by \Eq{Csigma:t}. \Eq{rev:4}
follows because the convex conjugate of $C_{\sigma_T}$ is
$R_{\sigma_T}(\vmu)=\ind\set{\vmu\in V_{\sigma_T}}+R(\vmu)$, and
$R_{\sigma_T}\bigParens{\vphi(\omega)}=R\bigParens{\vphi(\omega)}$
thanks to the compatibility of $\omega$ with $\sigma_T$. Finally,
\Eq{rev:5} follows from the definition of Bregman divergence, completing the proof.
\end{proof}

\section{Differentiability and controlled growth of $R$}
\label{app:growth}

The algorithm used by our market maker
requires a differentiable objective whose gradient does not grow too fast
as it approaches the boundary of $\M$. Note that for LMSR, the Bregman divergence is
formally not even differentiable in its first argument (it is subdifferentiable).
So, in addition to requiring the controlled growth of the gradient, we also need to assume that $R$ can be extended into a differentiable function.
Specifically, we say that $\bR:\R^\I\to(-\infty,\infty]$ is a \emph{convex extension} of $R$ if $\bR$ is convex and coincides with $R$ wherever $R<\infty$. We require existence of an extension with the controlled growth property in the following sense:
\begin{definition}
Let $\S\subseteq[0,1]^n$ be a compact convex set. We say
that a convex function $F$
exhibits \emph{controlled growth} on $\S$
if it is differentiable on $\S\cap(0,1)^n$ and
if there exists a fixed $p\ge 0$ and $L\ge 0$ such that
for any $\eps>0$, the gradient $\nabla F$ has a bounded
Lipschitz constant $L_\eps\le L\eps^{-p}$ over $\S\cap[\eps,1-\eps]^n$.
\end{definition}
\begin{assumption}%
\label{assume:growth}%
$R$ has a convex extension $\bR$ such that for all partial outcomes $\sigma$,
when $\bR$ is viewed as a function
on $\Vsigma$, it exhibits controlled
growth on $\M_\sigma$.
\end{assumption}
We write $\bR_\sigma$ for the restriction of $\bR$ to
$\Vsigma$. Note that this restriction is formally a function defined on a
space of dimension $\card{\I\wo\I_\sigma}$ and thus, formally, $\nabla\bR_\sigma$ has
the dimension $\card{\I\wo\I_\sigma}$. We extend $\nabla\bR_\sigma$ into a vector
in $\R^\I$ by inserting zeros at coordinates $i\in\I_\sigma$. A key consequence of
this construction is that for any partial outcome $\sigma$ and
all $\vmu\in\M_\sigma$ such that $\mu_i\in(0,1)$ for $i\not\in\I_\sigma$,
the gradient $\nabla\bR_\sigma(\vmu)$ is defined, and $\nabla\bR_\sigma(\vmu)\in\partial R_\sigma(\vmu)$. As a result, we have that $\vtheta=\nabla\bR_\sigma(\vmu)$ implies that $\nabla C_\sigma(\vtheta)=\vmu$ (but not vice versa). Assumption~\ref{assume:growth} can
be verified for instance by upper-bounding the operator norm of the Hessian, which directly
upper-bounds the Lipschitz constant of the gradient.
\begin{example}\emph{Controlled growth for LMSR.}
We define the extension of negative entropy over the non-negative orthant,
$\bR(\vmu)=\ind\set{\vmu\ge\vzero}+\sum_{i\in\I}\mu_i\log\mu_i$, which yields $\bR_\sigma(\vmu)=\ind\set{\mu_i\ge 0\text{ for all }i\not\in\I_\sigma}+\sum_{i\not\in\I_\sigma}\mu_i\log\mu_i$. The Hessian
is a diagonal matrix with entries $1/\mu_i$, so its operator norm is $\max_{i\not\in\I_\sigma} 1/\mu_i$, and thus $L_\eps=O(1/\eps)$, which satisfies the controlled growth condition with $p=1$.
\end{example}

\section{Proof of Proposition 4.1}
\label{app:proof:gap}

The guaranteed profit when moving from $\vtheta$ to $\hvtheta$ is
\begin{align}
\notag
&\min_{\omega\in\Omega}
  \BigBracks{ (\hvtheta-\vtheta)\inprod\vphi(\omega) - C(\hvtheta) + C(\vtheta)
  }
\\
\label{eq:gap:1}
&\qquad{}
= \min_{\vmu\in\M}
  \BigBracks{ (\hvtheta-\vtheta)\inprod\vmu - C(\hvtheta) + C(\vtheta)
  }
\\
\notag
&\qquad{}
= \min_{\vmu\in\M}
  \BigBracks{
    (\hvtheta-\vtheta)\inprod(\vmu-\hvmu)
    + \hvtheta\inprod\hvmu - C(\hvtheta)
    -  \vtheta\inprod\hvmu + C(\vtheta)
  }
\\
\label{eq:gap:2}
&\qquad{}
= \min_{\vmu\in\M}
  \BigBracks{
    (\hvtheta-\vtheta)\inprod(\vmu-\hvmu)
    + R(\hvmu)
    -  \vtheta\inprod\hvmu + C(\vtheta)
  }
\\
\label{eq:gap:3}
&\qquad{}
= \D{\hvmu}{\vtheta} - g(\hvmu)
\enspace.
\end{align}
\Eq{gap:1} follows because the minimized objective is linear
in $\vphi(\omega)$. \Eq{gap:2} follows from the definition of $R$.
Finally, \Eq{gap:3} follows because
$
  \nabla F(\hvmu)=\nabla \bR(\hvmu) - \vtheta
                 =\hvtheta-\vtheta
$,
and hence
\begin{equation}
\tag*{\qed}
  g(\hvmu)=\max_{\vmu\in\M}\BigBracks{(\hvtheta-\vtheta)\cdot(\hvmu-\vmu)}
\enspace.
\end{equation}
\renewcommand{\qed}{}%